\documentclass[12pt,a4]{article}
\usepackage{graphics}
\usepackage{graphicx}
\usepackage{amsfonts}
\usepackage{amssymb}
\usepackage{amsmath}
\usepackage{cite}
\usepackage{subfigure}

\newlength{\dinwidth}                                                      
\newlength{\dinmargin}                                                      
\def\lapproxeq{\lower .7ex\hbox{$\;\stackrel{\textstyle <}{\sim}\;$}}                                                      
\def\gapproxeq{\lower .7ex\hbox{$\;\stackrel{\textstyle >}{\sim}\;$}}

\def\bea{\begin{eqnarray}}                                                      
\def\eea{\end{eqnarray}}                                                      
\def\Ktbold{\mbox{\boldmath${k}$}_t} 
\def\qbold{\mbox{\boldmath${q}$}}

\begin{document}                                                        
\titlepage    
\vspace*{1in}    
\begin{center}    
{\Large \bf Unintegrated parton distributions of pions and nucleons from the CCFM equations in the single-loop 
approximation}   \\
\vspace*{0.4in}    
Agnieszka Gawron, Jan Kwieci\'nski, and Wojciech Broniowski 
  
\vspace*{0.5cm}    
  
{\it The H. Niewodnicza\'nski Institute of Nuclear Physics, PL-31342 Cracow, Poland} \\    
\end{center}    
\vspace*{1cm}    
    
\vskip1cm    

\begin{abstract}

The unintegrated quark and gluon distributions in the pion and nucleon 
are analysed using the CCFM equations in the single-loop approximation.   
We utilise the transverse-coordinate (or impact-parameter) representation 
which diagonalises the equations and study in detail the
dependence on the transverse coordinate induced by the CCFM evolution. 
We find considerable broadening 
of the transverse-momentum distributions with an increasing magnitude of the hard scale, $Q$.
For instance, at the Bjorken $x=0.1$ the root mean squared transverse momentum of the gluons is enhanced by about 
1~GeV when evolved from the the initial low scale to $Q^2=10~{\rm GeV}^2$, and by about 2~GeV when evolved 
up to to $Q^2=100~{\rm GeV}^2$.
The broadening effect is enhanced with decreasing Bjorken $x$, and is stronger for the gluons than for 
the quarks. 
Analytic solution for the average transverse momentum 
corresponding to the $x$-moments of the distributions is obtained. 
The parton luminosities are also discussed.       
\end{abstract}
                                                   
\section{Introduction}
Parton distributions are fundamental and universal quantities of the QCD-improved parton model. 
They depend on the longitudinal momentum fraction of the parent hadron, $x$, 
as well as on the 
hard scale of the process, $Q$. The inclusive cross-sections describing  hard processes are   governed 
by the (collinear) factorisation theorem expressing those cross-sections as convolutions of partonic
distributions and hard (partonic) cross-sections \cite{ESW}. On the other hand, less inclusive 
measurements are often sensitive to the transverse momenta of the partons and require an
introduction of the {\em unintegrated} parton distributions, {\em i.e.} distributions which are 
not integrated over the transverse momentum of the parton 
\cite{DDT,KMR1,KMKS1,KMR2,MR,KHMR,MIU,SZCZUREK1,LUNDSMX,COLLINS}.  
These  unintegrated distributions depend then upon the transverse momentum of the parton 
besides the dependence on  $x$ and  $Q$. 
Typical  cases 
where the information on the transverse momentum of partons is needed are the prompt 
photon production, or the transverse momentum distributions of the Drell-Yan lepton pairs, 
$W^{\pm}$ and $Z^0$ bosons, {\em etc.}  The unintegrated parton distributions are also needed 
in the calculations made within the $k_T$-factorisation framework 
\cite{KTFAC,SZCZUREK2,MOTYKA,ZOTOV,IVANOV}.

The unintegrated parton distributions are described by the Catani-Ciafaloni-Fiorani-
Mar\-che\-si\-ni (CCFM) equations 
\cite{CCFM,GM1,BRW,GMBRW,GM2,KMSU,CCFMD,KGBGT,GSAL,JUNG,JUNGS} which are based on the quantum coherence 
implying angular ordering along the partonic cascade \cite{DKTM}.  
In the so-called single-loop 
approximation \cite{BRW,GMBRW}, adequate for large and moderately small values of 
$x$ (say, $x > 0.01$), the CCFM equations reproduce the conventional leading-order DGLAP equations for the 
integrated distributions. Thus, they may be viewed as an extension of the DGLAP evolution 
which allows to investigate more general quantities.

The purpose of this paper is to analyse in detail the unintegrated parton distributions 
of pions and nucleons.  To this aim we explore the system of the CCFM equations 
in the single loop 
approximation utilising the transverse-coordinate representation of the 
parton distributions. The 
transverse-coordinate, denoted in this paper by $b$, is a variable conjugate 
to the transverse momentum of the parton. It
has been widely used in the study of the soft gluon resummation effects in the
$e^{+}e^{-}$ collisions \cite{BASSETTO,KODAIRA}, in the  
transverse-momentum distributions of the Drell-Yan pairs or gauge bosons \cite{PT},  
{\em etc.}  The formalism 
adopted in our analysis is similar to methods used in those studies. An important merit of 
our use of the transverse-coordinate representation is the fact that it {\em diagonalises} the 
system of the CCFM equations in the single-loop approximation \cite{JK,AGJK}.  
In practice, this means that the equations are solved independently for each value of $b$.
Also, in this approximation the  
non-perturbative effects encoded in the choice of the initial dependence on $b$
at the  reference scale $Q_0$ (the profile function)
factorise in the solution at a scale $Q$.
This input profile  introduces the damping factor for large values of $b$ 
 that makes  it   possible to extend 
smoothly  the partonic $k_T$-distributions down to the point $k_T=0$.    
The CCFM evolution changes   
the dependence on $b$ and leads to significant broadening of the transverse momentum distributions.          
The average transverse momentum squared grows approximately linearly with $Q^2$ 
(modulo logarithmic corrections).  
The effect is strongest for gluons, when e.g. for $Q^2=100~{\rm GeV}^2$  
the CCFM evolution is found to generate   
  $\langle k_T^2 \rangle_{\rm evol}^{1/2}$  
about $2.3~{\rm GeV}$ for $x=0.1$ and  $3.2~{\rm GeV}$ for $x=0.01$. These numbers 
should be compared to the initial spreading, $\langle k_T^2 \rangle_F$, which is 
typically assumed to be in the range of $(0.5-1~{\rm GeV})$. 
   Broadening of the 
$k_T$ distribution corresponds to the development of the long-range 
tail in the distribution at large $k_T$. \\

The average transverse momenta 
squared can be expressed in terms  of the logarithmic derivative of the distributions in the $b$ 
representation at $b=0$.  
Factorisation of the input profile implies that the average transverse momentum squared 
of the parton is the sum of the 'primordial' (non-perturbative) component at $Q_0$ 
and the term generated by the evolution. The unintegrated parton 
distributions are  constrained  by the integrated distributions  which correspond to the 
integrals of the unintegrated 
distributions over the transverse momentum squared of the partons.  
In order to make 
our analysis realistic we use constraints implied  by the results of the global 
LO QCD analysis \cite{GRS,GRV} for the integrated  distributions of pions and nucleons.    
To be precise, we take the same parametrisation of the 
starting integrated distributions as those used in \cite{GRS,GRV} and assume,
 as an educated guess,
a Gaussian transverse momentum distribution, uniform for gluons and quarks
\footnote{The assumption of uniformity may be lifted with no difficulty, as well as other initial
parameterisations than \cite{GRS,GRV} may be tested. This has little influence on our general 
conclusion on the large dynamical spreading of the distributions.}. Combining this 
assumption with the evolution from the CCFM equations in the single-loop approximation  
one should get a set of realistic unintegrated parton distributions   
hadrons. These can  be used in phenomenological applications in the  mentioned region 
of large and moderately small values of $x$, where the single-loop approximation 
 is expected to be adequate.

The content of our paper is as follows: 
In the next section we recall the system of the CCFM equations in the single-loop approximation 
and discuss its transverse coordinate representation.  In Sec. 3  we discuss
the unintegrated parton distributions of pions and nucleons which follow from the 
numerical solution of the CCFM equations 
in the single-loop approximation.  We  present our results for the $b$ profiles and for     
the transverse-momentum distributions, as well as discuss the increase of the average transverse 
momentum with the increasing magnitude of the hard scale, $Q$, at different 
values of $x$.  In Sec. 4 we discuss  
the moments of the $b$ profiles and present results for the average transverse momenta squared,
$<k_T^2(n,Q^2)>_i$, for the $n$th moment for partons of species $i$.  
We also derive (semi)analytic expressions for $<k_T^2(n,Q^2)>_i$.  
In Sec. 5 we discuss partonic 
luminosities and, finally, in Sec. 6 we give a summary of our 
results. 
         
\section{CCFM equations in the single loop approximation}
  The  original Catani-Ciafaloni-Fiorani-Marchesini
(CCFM) equation \cite{CCFM,GM1} for the unintegrated, scale-dependent gluon distribution
$f_G(x,k_T,Q)$, which is generated by the sum of ladder diagrams with  angular
ordering along the chain,  has the following form:
\begin{eqnarray}
f_G(x,k_T,Q)&=& \tilde f_G^0(x,k_T,Q) + \int{d^2\qbold \over \pi q^2}
\int_x^1 {dz\over z} \Theta(Q-qz) \Theta(q-q_0) { \alpha_s\over 2\pi } \Delta_S (Q,q,z) \nonumber \\
&\times& \left[2N_c\Delta_{NS}(k_T,q,z) + {2N_cz\over (1-z)} f\left({x\over z},|\Ktbold+(1-z)\qbold|,q\right)\right] ,
\label{ccfm1}
\end{eqnarray}
where $\Delta _S(Q,q,z)$ and $\Delta_{NS}(k_T,q,z)$ are the Sudakov and non-Sudakov
form factors,  
\begin{eqnarray}
\Delta _S(Q,q,z)&=& \exp\left[-\int_{(qz)^2}^{Q^2}{dp^2\over
p^2}{\alpha_s\over 2 \pi} \int _0^{1-q_0/p}dzzP_{gg}(z)\right],
\label{ds} \\
\Delta_{NS}(k_T,q,z)&=& \exp\left[-\int_z^1{dz'\over z'}
\int_{(qz')^2}^{k_T^2} {dp^2\over p^2}{2N_c\alpha_s\over 2 \pi}\right] .
\label{dns}
\end{eqnarray}
The variables $x$, $k_T$, and $Q$ denote the longitudinal momentum fraction,
the transverse momentum of the gluon, and the hard scale, respectively. The
latter is defined in terms of the maximum emission angle \cite{LUNDSMX,CCFM}.
The constraint $\Theta(Q-qz)$ in Eq.~(\ref{ccfm1}) reflects the angular ordering, and the inhomogeneous term,
$\tilde f^0(x,k_T,Q)$, is related to the input non-perturbative gluon distribution.
It also contains effects of both the Sudakov and non-Sudakov form-factors \cite{KMSU}.

Equation (\ref{ccfm1})  in a sense interpolates between 
a fragment of 
the DGLAP evolution at large $x$ 
and the BFKL dynamics at small $x$.  To be precise, it contains only the $g\rightarrow gg$ 
splittings and only those  parts of the 
 splitting function $P_{gg}(z)$ which are singular at either $z \rightarrow 1$ 
or $z \rightarrow 0$. In order to make the CCFM framework more 
 realistic in the region of large 
and moderately small values of $x$ we make use of the following extension 
\cite{AGJK}: 
\begin{enumerate}
\item We introduce, besides the unintegrated gluon distributions $f_G(x,k_T,Q)$,
also the unintegrated quark and antiquark gluon distributions, $f_{q_i}(x,k_T,Q)$ and  
 $f_{\bar q_i}(x,k_T,Q)$.  
\item We include, in addition to the $g\rightarrow gg$ splittings, also the the $q\rightarrow gq$, 
$\bar q \rightarrow g\bar q$, and $g \rightarrow 
\bar q q$ transitions along the chain.  
\item We take into account the complete splitting functions $P_{ab}(z)$, and not only their singular 
parts.   
\end{enumerate}
In the region of large and moderately small values of the parameter $x$ ($x \ge 0.01$ or so) 
one can  introduce the single-loop approximation that corresponds to the following 
replacements: 
\begin{eqnarray}
\Theta(Q-qz) &\rightarrow& \Theta(Q-q),
\label{sloop1}
\nonumber \\
\Delta_{NS}(k_T,q,z) &\rightarrow& 1. \label{sloop2}
\end{eqnarray}     
It is also useful to 'unfold' the Sudakov form factor such that  
the real emission and virtual terms appear on an equal footing in the corresponding 
evolution equations.  Finally, the unfolded system of the CCFM equations in the single-loop 
approximation has the following form: 
\begin{eqnarray}
f_{NS}^i(x,k_T,Q)&=& f^{i0}_{NS}(x,k_T) +\int_0^1dz\int{d^2q\over \pi q^2} {\alpha_s(q^2)\over
2\pi}\Theta(q^2-q_0^2)\Theta(Q-q) P_{qq}(z) \\
&\times& \left[\Theta(z-x)f_{NS}^i\left({x\over z},k^{\prime}_T,q\right)-f_{NS}^i(x,k_T,q) \right] ,
\label{ccfmns} \nonumber \\
f_{S}(x,k_T,Q)&=&f_S^0(x,k_T)+ \int_0^1dz\int{d^2q\over \pi
q^2} {\alpha_s(q^2)\over 2\pi}\Theta(q^2-q_0^2)\Theta(Q-q)  \\
&\times& \bigg\{\Theta(z-x)\left[P_{qq}(z)f_{S}\left({x\over
z},k^{\prime}_T,q\right)+ P_{qg}(z)f_{G}\left({x\over
z},k^{\prime}_T,q\right)\right]-P_{qq}(z)f_{S}(x,k_T,q)\bigg\}, \nonumber \\
f_{G}(x,k_T,Q)&=&f_G^0(x,k_T)
+ \int_0^1dz\int{d^2q\over \pi q^2} {\alpha_s(q^2)\over
2\pi}\Theta(q^2-q_0^2)\Theta(Q-q) \\
&\times& \bigg\{\Theta(z-x)\left[P_{gq}(z)f_{S}\left({x\over
z},Q^{\prime}_t,q\right)+ P_{gg}(z)f_{G}\left({x\over
z},Q^{\prime}_t,q\right)\right] \bigg. \nonumber \\ 
&& - \bigg. \big[zP_{gg}(z)+zP_{qg}(z)\big]f_{G}(x,k_T,q)\bigg\} , \nonumber
\label{ccfmg}
\end{eqnarray}
where
\begin{equation}
{\bf k^{\prime}_T} ={\bf k_T} + (1-z){\bf q} .
\label{qprimet}
\end{equation}
The functions $f_{NS}^i(x,k_T,Q)$ are the unintegrated non-singlet quark  distributions, while the 
unintegrated singlet distribution, $f_{S}(x,k_T,Q)$, is defined as 
\begin{equation}
f_S(x,k_T,Q)=\Sigma_{i=1}^f [f_{q_i}(x,k_T,Q)+ f_{\bar q_i}(x,k_T,Q)]
\label{singlet}
\end{equation}
The functions $P_{ab}(z)$ are the LO splitting functions corresponding to real 
emissions, {\em i.e.}:
\begin{eqnarray}
P_{qq}(z)&=&{4\over 3} {1+z^2\over 1-z} ,\\
P_{qg}(z)&=&N_f[z^2+(1-z)^2] ,\\
P_{gq}(z)&=&{4\over 3} {1+(1-z)^2\over z} ,\\
P_{gg}(z)&=&2N_c\left[{z\over 1-z}+{1-z\over z} + z(1-z)\right] .
\label{splitf}
\end{eqnarray}
where $N_f$ and $N_c$ denote the number of flavours and colours, respectively. 
After integrating over $d^2{\bf k_T}$ on both sides of 
Eqs.~(\ref{ccfmns} - \ref{ccfmg}) we get the usual DGLAP equations for the 
integrated parton distributions $p_k(x,Q^2)$, defined by
\begin{equation}
xp_k(x,Q^2)=\int_0^\infty dk_T^2 f_k(x,k_T,Q).
\label{idis}
\end{equation}

Equations
(\ref{ccfmns})-(\ref{ccfmg}) can be diagonalised by the Fourier-Bessel transform,
\begin{eqnarray}
f_{k}(x,k_T,Q)&=&\int_0^{\infty}db b J_0(k_Tb)\bar f_{k}(x,b,Q) ,\label{fb1}\\
\bar f_{k}(x,b,Q)&=&\int_0^{\infty}dk_T k_T J_0(k_Tb) f_{k}(x,k_T,Q) ,
\label{fb2}
\end{eqnarray}
where $k=NS$, $S$, or $g$,  and $J_0(u)$ is the Bessel function. At $b=0$ the 
functions $\bar f_{k}(x,b,Q)$  are related to the integrated distributions $p_i(x,Q^2)$,
\begin{equation}
\bar f_{k}(x,b=0,Q)={1\over 2}xp_k(x,Q^2).
\label{fbxp}
\end{equation} 
The corresponding 
evolution  equations for $\bar f_{NS}(x,b,Q)$, $\bar f_{S}(x,b,Q)$, and $\bar f_{g}(x,b,Q)$,
equivalent to Eqs.~(\ref{ccfmns}) - (\ref{ccfmg}) read
\begin{eqnarray}
&&Q^2{\partial \bar f_{NS}(x,b,Q)\over \partial Q^2}= \nonumber \\
&&{\alpha_s(Q^2)\over 2\pi}
\int_0^1dz 
P_{qq}(z) 
\left[\Theta(z-x)J_0[(1-z)Qb]\bar f_{NS}\left({x\over z},b,Q\right)-zP_{qq}(z)\bar f_{NS}(x,b,Q)
\right],
\label{dccfmnsb} \\
&& Q^2{\partial \bar f_{S}(x,b,Q)\over \partial Q^2}= \nonumber \\ 
&&{\alpha_s(Q^2)\over 2\pi} \int_0^1dz
\bigg\{\Theta(z-x)J_0[(1-z)Qb]\left[P_{qq}(z)\bar f_{S}\left({x\over z},b,Q\right)+ 
P_{qg}(z)\bar f_{G}\left({x\over z},b,Q\right)\right] \bigg. \nonumber \\
&& \bigg.-[zP_{qq}(z)+zP_{gq}(z)]\bar f_{S}(x,b,Q)\bigg\},
\label{dccfmsb} \\
&&Q^2 { \partial \bar f_{G}(x,b,Q)\over \partial Q^2}= \nonumber \\  
&&{\alpha_s(Q^2)\over 2\pi}\int_0^1dz 
\bigg\{\Theta(z-x)J_0[(1-z)Qb]\left[P_{gq}(z)\bar f_{S}\left({x\over z},b,Q\right)+ 
P_{gg}(z)\bar f_{G}\left({x\over z},b,Q\right)\right] \bigg. \nonumber \\ 
&&\bigg. -[zP_{gg}(z)+zP_{qg}(z)]\bar f_{G}(x,b,Q)\bigg\},
\label{dccfmgb}
\end{eqnarray}
with the initial conditions
\begin{equation}
\bar f_k(x,b,Q_0)=\bar f_k^0(x,b).
\label{bcond}
\end{equation}

In our analysis of the CCFM equations we assume,  
for simplicity and from the lack of detailed experimental knowledge,  a factorisable form 
of the initial conditions (\ref{bcond})
\begin{equation}
\bar f_k^0(x,b)={1\over 2}F(b)xp_k(x,Q_0^2). 
\label{bcondf}
\end{equation}
Other, more complicated and non-factorisable 
initial conditions may be explored with no difficulty as well.
The input profile function, $F(b)$, is linked through the Fourier-Bessel transform 
(\ref{fb2}) to the 
non-perturbative $k_T$ distribution at the scale $Q_0$.  At $b=0$ we have 
the normalisation condition for the profile function, $F(0)=1$.

\section{Unintegrated parton distributions in the pion and nucleon}

Since the initial condition for the QCD evolution, Eq. (\ref{bcondf}), assumes a uniform 
dependence on the profile function $F(b)$ for all distribution functions $f_i$,
it is convenient to factorise this function by introducing the 'tilde' distributions,
\begin{eqnarray}
\tilde{f}_i(x,b,Q)=\bar {f}_i(x,b,Q)/F(b), \;\;\;\;\;\; i={\rm G, S, NS}. \label{rescaled}
\end{eqnarray}
The evolution equations (\ref{dccfmnsb}-\ref{dccfmgb}) can then be written in terms of the $\tilde{f}$ functions 
only. In other words, all dynamical information on the $b$-dependence, generated by the evolution and linked to
the $Q$ and $x$ variables, is contained in the $\tilde{f}$ functions, while the uniform initial profile $F(b)$
is carried over as a multiplicative factor  and is decoupled from the dynamics. 
Certainly, we may choose different profiles $F(b)$, which affects
directly the shape of the unintegrated distribution functions, nevertheless the evolution is independent 
of this choice,  as long as the factorisation is assumed. 
For this reason we first present the numerical results for the scaled functions 
$\tilde{f}_i(x,b,Q)$.

The evolution equations for the $\tilde{f}$ functions are solved numerically.  
The method used  is based on the
discretisation made with the help of the Chebyshev polynomials  and is an 
extension of the method developed in \cite{JKDSK} for the solution of the DGLAP 
equations.  \\

\begin{figure}[b]
\begin{center}
\includegraphics[width=13.5cm]{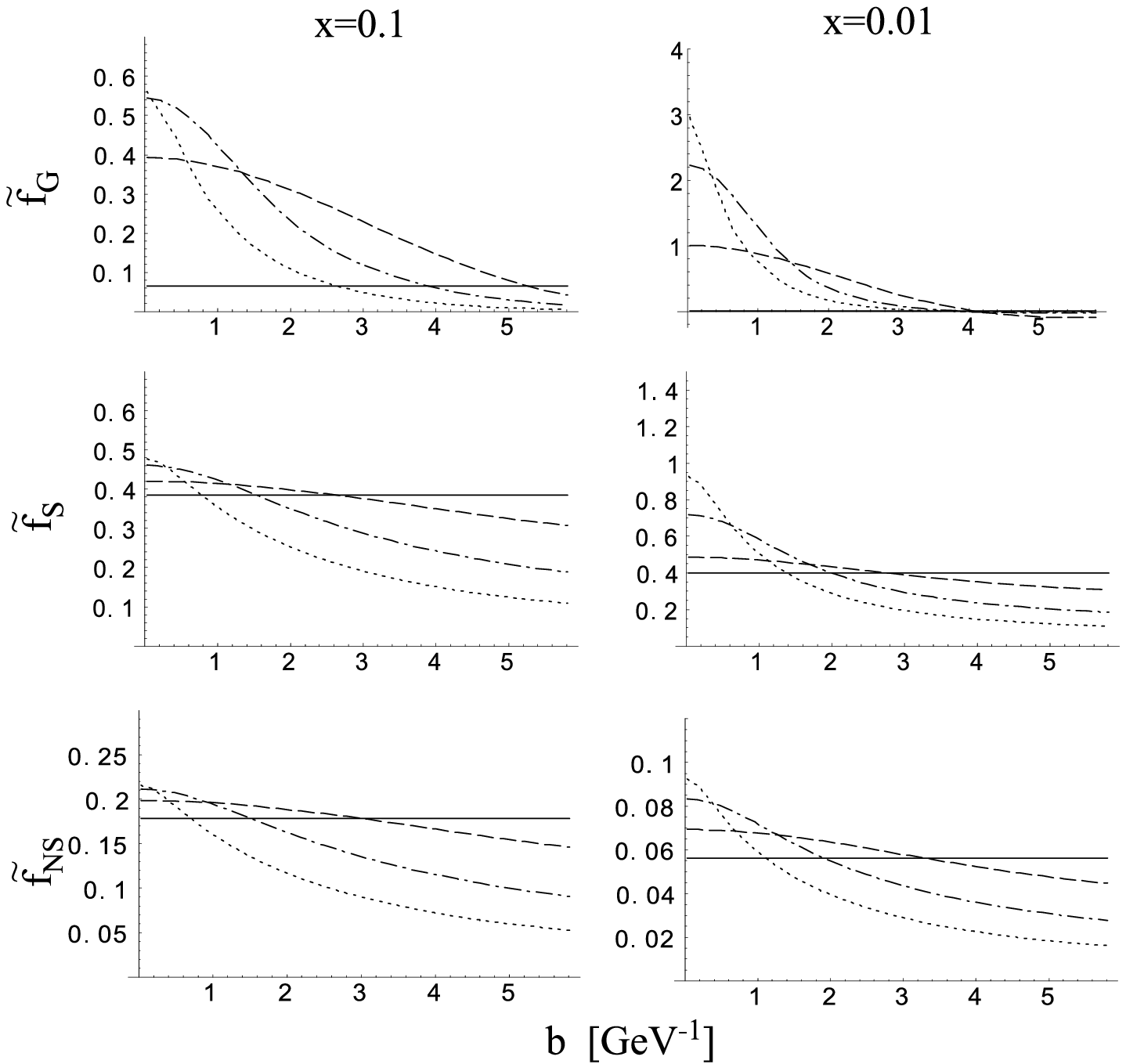}
\end{center}
\caption{Solutions of the CCFM equations in the single-loop 
approximation, Eqs.~(\ref{dccfmnsb}-\ref{dccfmgb}), for the 
unintegrated partonic distribution functions in the pion. The gluon, $\tilde{f}_G$, quark singlet, 
$\tilde{f}_S$, and quark non-singlet distributions, $\tilde{f}_{NS}$, are plotted as 
functions of the transverse coordinate, $b$. The left and right parts are for $x=0.1$ and $x=0.01$, respectively.
The solid lines correspond to the initial scale of the GRS parameterisation, $Q^2=Q_0^2=0.26~{\rm GeV}^2$ 
\cite{GRS}, while the dashed, dash-dotted and dotted lines correspond to $Q^2=1~{\rm GeV}^2$, 
$10~{\rm GeV}^2$, and $100~{\rm GeV}^2$, respectively.  } \label{fig:bGRS}
\end{figure}

Figure \ref{fig:bGRS} shows the solutions of Eqs.~(\ref{dccfmnsb}-\ref{dccfmgb}) for the case of the pion 
with the LO Gl\"uck-Reya-Schienbein (GRS) distributions \cite{GRS} taken at the initial scale 
$Q_0^2=0.26~{\rm GeV}^2$.
The left and right sides of the figure display the results for $x=0.1$ and $x=0.01$, {\em i.e.} moderately
large and moderately small values of the Bjorken variable. We show, from top to bottom, the 
distributions $\tilde{f}_G$, $\tilde{f}_S$, and $\tilde{f}_{NS}$, plotted as 
functions of the transverse coordinate, $b$. The various types of lines
indicate consecutive values of the momentum $Q$, with the solid line corresponding to the initial scale 
of the GRS parameterisation, $Q^2=Q_0^2=0.26~{\rm GeV}^2$ 
\cite{GRS}, and the dashed, dash-dotted and dotted lines denoting results 
for $Q^2=1~{\rm GeV}^2$, 
$10~{\rm GeV}^2$, and $100~{\rm GeV}^2$, respectively. The original
distributions at $Q=Q_0$ are flat in $b$, since we use the rescaled functions (\ref{rescaled}).
We note immediately from the figure that the increase of $Q$ 
results in sharp peaking of the distributions at $b=0$, with the effect stronger at the smaller value of $x$.
The shrinkage of the distributions in $b$ is most prominent for the gluons. For instance, for the gluons at $x=0.01$ 
the curves drop to half of the values at $b=0$ around $b=2$, $1.2$, and $0.6~{\rm GeV}^{-1}$ for $Q^2=1$, $10$, and 
$100~{\rm GeV}^2$, respectively. To the extent these distributions can be approximated with a Gaussian, that would
correspond to widths in the transverse momentum of about $0.8~{\rm GeV}$, $1.4~{\rm GeV}$,  
and $3~{\rm GeV}$,  respectively. The transverse-momentum distributions will be discussed in detail 
in the following parts of the paper. At the values of $x$ displayed in Fig. \ref{fig:bGRS}
the shrinking effect in $b$ is similar for the singlet and non-singlet quark distributions.

We note that for the case of gluons at $x=0.01$ the function $\tilde{f}_G$ becomes negative above 
$b\sim 4~{\rm GeV}^{-1}$. To our knowledge, this poses no physical problems. Positivity is required 
for the unintegrated distributions {\em in the transverse-momentum space},  since then
the positivity of cross-sections, which are physical 
quantities, is guaranteed. The Fourier-Bessel transform of a 
positive function, however, need not be positive itself. We will see shortly that in the 
transverse-momentum space the distributions are 
positive. The effect of negative $\tilde{f}_G$ at larger $b$ and small $x$ is an artifact of the valence-like 
input for the gluons in the parameterisations of Refs.~\cite{GRS,GRV}, and 
can be understood as follows: At small $x$ the evolution equation 
(\ref{dccfmgb}) is dominated on the right-hand side by the gluon contribution, hence 
can be approximated with the following form:  
\begin{eqnarray}
&& Q^2 { \partial \tilde f_{G}(x,b,Q)\over \partial Q^2}=  \nonumber \\
&&{\alpha_s(Q^2)\over 2\pi}\int_0^1dz 
J_0[(1-z)Qb]\left\{P_{gg}(z)\tilde f_{G}\left({x\over z},b,Q\right)-
[zP_{gg}(z)+zP_{qg}(z)]\tilde f_{G}(x,b,Q)\right\} \nonumber \\
&&- \{zP_{qg}(z)+zP_{gg}(z)[1-J_0((1-z)Qb)]\tilde f_{G}(x,b,Q)\}
\label{ccfmbsx}
\end{eqnarray}
At small $x$ the dominant region of integration in the first term on the right-hand 
side of Eq.~(\ref{ccfmbsx}) is $x<z<<1$, hence we may approximate $1-z$ 
by $1$ in the argument of the Bessel function in this term.  
The solution of Eq.~(\ref{ccfmbsx}) with the initial condition
\begin{equation}
\tilde f_{G}(x,b,Q_0)={1\over 2} xg(x,Q_0^2)
\label{bctild}
\end{equation}
 then reads 
\begin{equation}
\tilde f_{G}(x,b,Q_0)={1\over 2}T_g(b,Q)xg(x,Q_{\rm eff}^2(b,Q)),
\label{soltil}
\end{equation}
where
\begin{equation}
T_g(b,Q)=\exp\left\{\int_{Q_0^2}^{Q^2}{dq^2\over q^2}{\alpha_s(q^2)\over 2\pi}\int_0^1
dz \left [zP_{qg}(z)+zP_{gg}(z)[1-J_0((1-z)qb)] \right ]\right\}
\label{tg}
\end{equation}
and $Q_{\rm eff}^2(b,Q)$ is obtained from the following implicit equation:  
\begin{equation}
\int_{Q_0^2}^{Q_{\rm eff}^2(b,Q)}{dq^2\over q^2}{\alpha_s(q^2)\over 2 \pi}=
\int_{Q_0^2}^{Q^2}{dq^2\over q^2}{\alpha_s(q^2)\over 2 \pi}J_0(qb).
\label{qeff2}
\end{equation}
The function $xg(x,Q^2)$ is the integrated gluon distribution corresponding to the 
solution of equation (\ref{ccfmbsx}) at $b=0$ with the initial condition 
$xg(x,Q_0^2)$.  At sufficiently large $b$ the integral on the right-hand side of 
Eq.~(\ref{qeff2}) becomes negative, which gives $Q_{\rm eff}^2(b,Q)<Q_0^2$, and   
the effective 'backward'   
evolution of the valence gluons  gives a negative $xg(x,Q_{\rm eff}^2)$ for sufficiently small $x$ 
and for $Q_{\rm eff}^2(b,Q)$ sufficiently smaller than $Q_0^2$.  This is the technical
reason for the behaviour seen in Fig.~\ref{fig:bGRS}. 

\begin{figure}[b]
\begin{center}
\includegraphics[width=9cm]{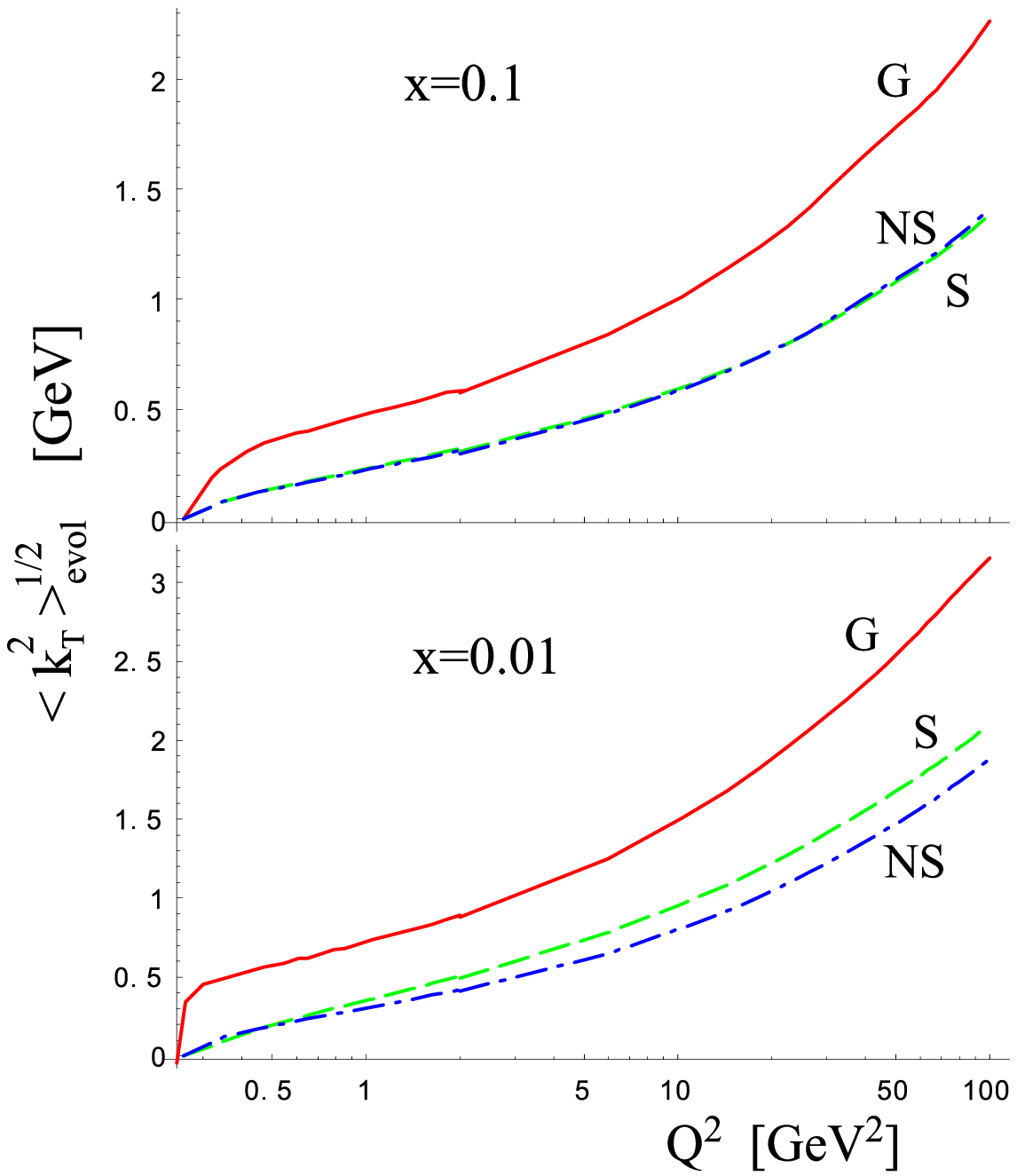}
\end{center}
\caption{The dynamically-generated root mean-squared transverse momentum of unintegrated partonic distribution 
functions in the pion, plotted as a function of $Q^2$. The top and bottom parts correspond to 
x=0.1 and x=0.01, respectively. The solid, dashed, and dot-dashed lines denote gluons, singlet quarks, and 
non-singlet quarks. The CCFM equations in the single-loop approximation with the initial GRS condition 
\cite{GRS} at the scale 
$Q^2=Q_0^2=0.26~{\rm GeV}^2$ are used for the evolution.}
\label{fig:ktGRS}
\end{figure}

The mean squared transverse momentum for a given distribution is equal to 
\begin{eqnarray}
\langle k_T^2(x,Q^2) \rangle_i \equiv \frac{\int d^2 k_T \, k_T^2 f_i(x,k_T,Q)}
{\int d^2 k_T \, f_i(x,k_T,Q)}=
-4 \left . \frac{d/db^2 \, f_i(b,x,Q)}{f_i(b,x,Q)} \right |_{b=0},
\end{eqnarray}
where the last equality follows from the expansion $J_0(z)=1-\frac{1}{4}z^2 + \dots$. Through the use of the
decomposition (\ref{rescaled}) we may immediately write 
\begin{eqnarray}
\langle k_T^2(x,Q^2) \rangle_i&=&\langle k_T^2 \rangle_F+\langle k_T^2(x,Q^2) \rangle_{\rm i, evol}, \\
\langle k_T^2 \rangle_F&=&-4 \left . \frac{d/db^2 \, F(b)}{F(b)} \right |_{b=0}, \nonumber \\
\langle k_T^2(x,Q^2) \rangle_{\rm i, evol}&=&-4 \left . \frac{d/db^2 \, \tilde{f}_i(b,x,Q)}{\tilde{f}(b,x,Q)} \right |_{b=0},
\end{eqnarray}
which means that the total mean squared transverse momentum is the sum 
of the contribution from the profile $F$, which is 
constant and corresponds to the width at the initial scale $Q=Q_0$, and the piece 
$\langle k_T^2 \rangle_{\rm i, evol}$, which is due entirely to the evolution and is 
independent of the profile $F$.
Because of these features we concentrate on $\langle k_T^2 \rangle_{\rm evol}$ in the discussion below.

In Fig.~\ref{fig:ktGRS} we show $\langle k_T^2 \rangle^{1/2}_{\rm evol}$, plotted as a function of $Q^2$,  
for $x$=0.1 (top) and $x$=0.01 (bottom). The solid, dashed, and dot-dashed lines denote gluons, singlet, and 
non-singlet quarks, respectively. Again, we clearly note the spreading effect with increasing $Q^2$. 
The effect is strongest for gluons, which at 
$Q^2=100~{\rm GeV}^2$ achieve $\langle k_T^2 \rangle_{\rm evol}^{1/2}$ 
about $2.3~{\rm GeV}$ for $x=0.1$ and  $3.2~{\rm GeV}$ for $x=0.01$. However, even at moderate 
$Q^2$ the spreading effect is sizeable, {\em e.g.} at $Q^2=10~{\rm GeV}^2$ and $x=0.01$ we find
$\langle k_T^2 \rangle^{1/2}_{\rm evol}=1.5~{\rm GeV}$, $1.0~{\rm GeV}$, and $0.8~{\rm GeV}$ 
for gluons, singlet, and non-singlet quarks, respectively.
The quoted numbers should be compared to the initial spreading, $\langle k_T^2 \rangle_F$, which is 
typically assumed to be in the range of $(0.5-1~{\rm GeV})^2$. 
We also note from Fig. (\ref{fig:ktGRS}) that at larger values of $x$ the quark singlet and non-singlet
widths are practically equal, and at lower $x$ the singlet distribution is 
somewhat wider. 

We also note a very sharp growth of $\langle k_T^2 \rangle^{1/2}_{\rm G,evol}$ at low $Q$ and low values of $x$. 
This feature  is an artifact of the valence-like gluon distributions 
at the reference scale $Q_0^2$.   

\begin{figure}[b]
\begin{center}
\subfigure{\includegraphics[angle=-90,width=0.43\textwidth]{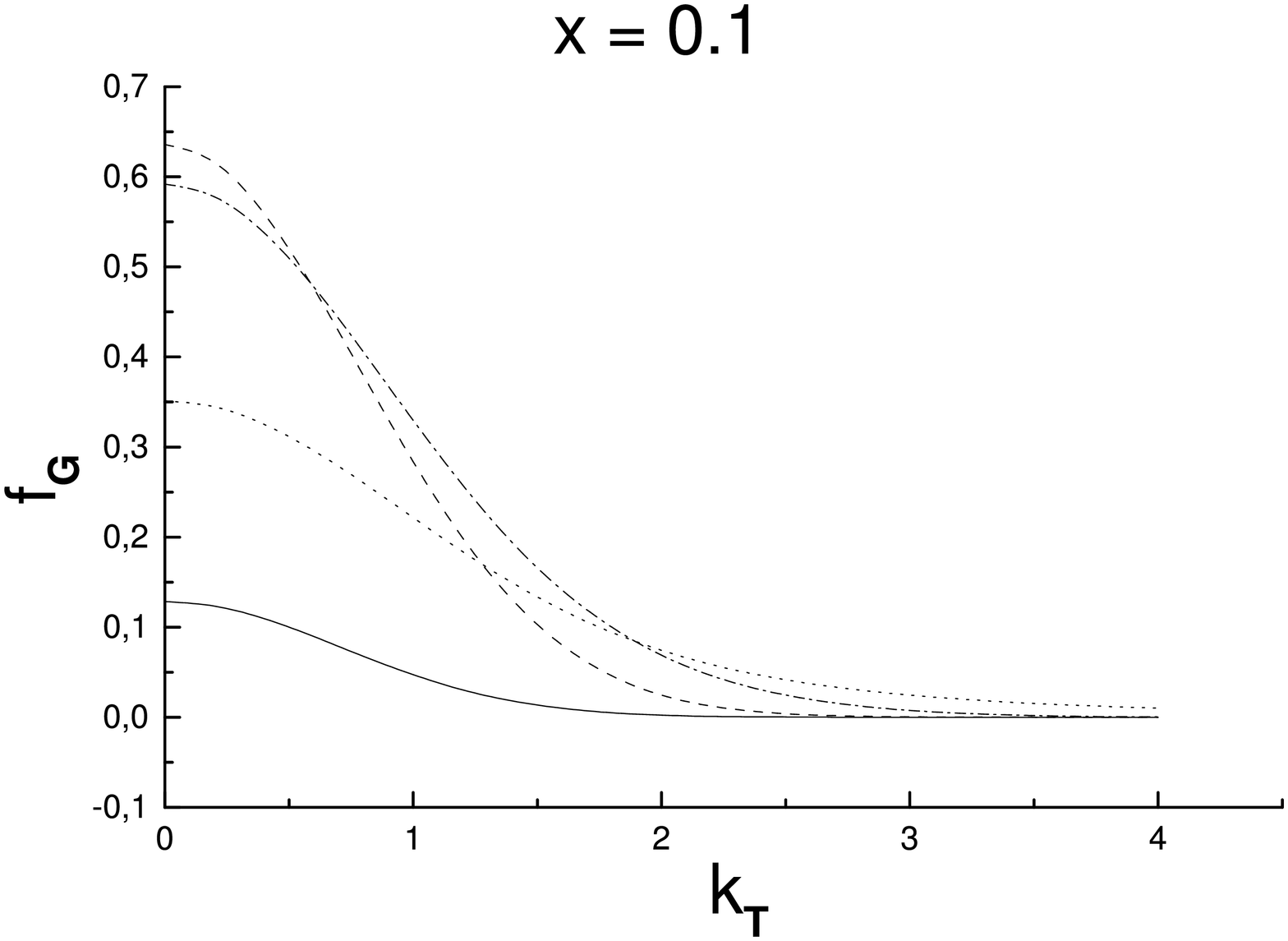}}
\subfigure{\includegraphics[angle=-90,width=0.43\textwidth]{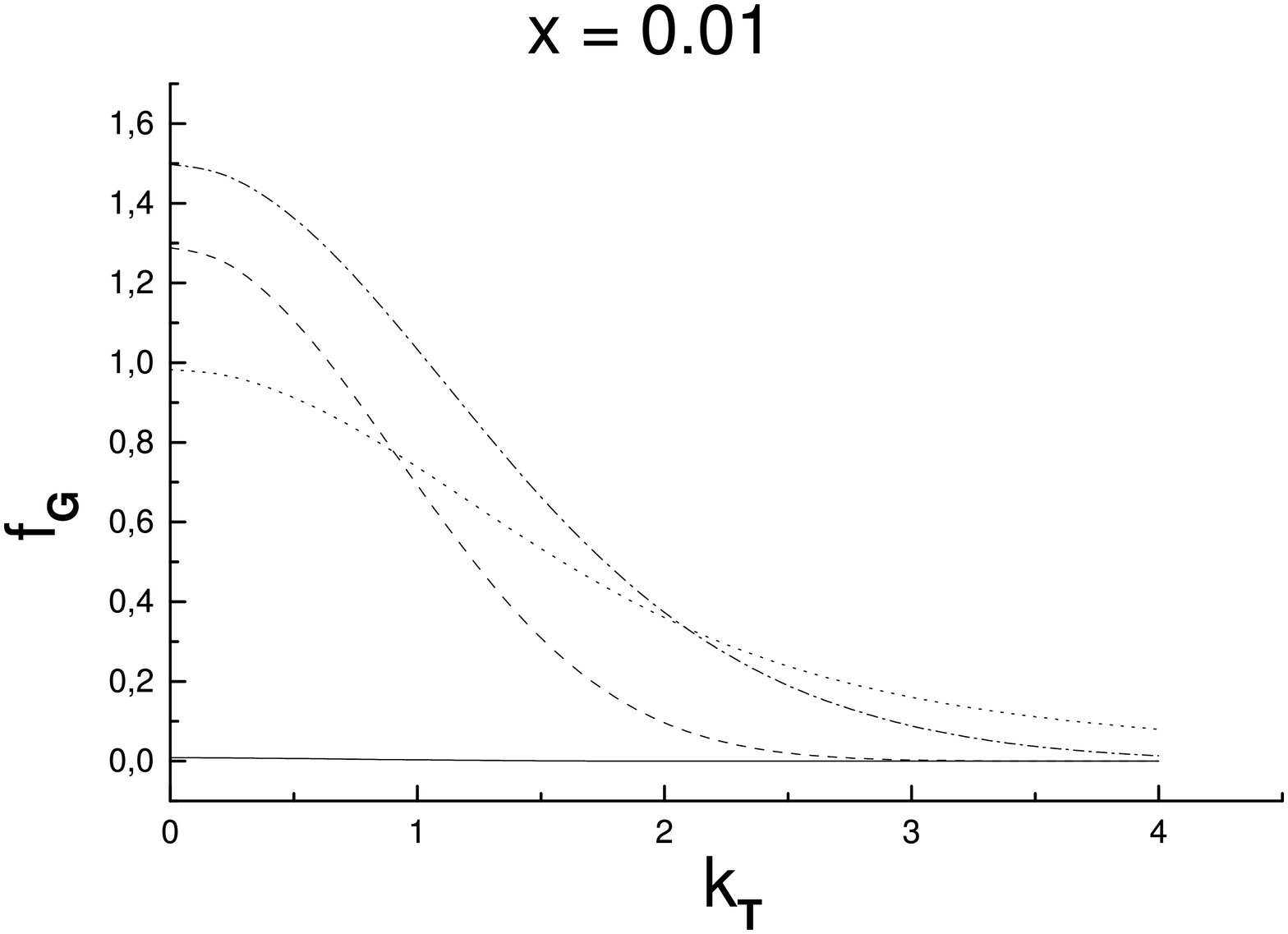}}\\
\subfigure{\includegraphics[angle=-90,width=0.43\textwidth]{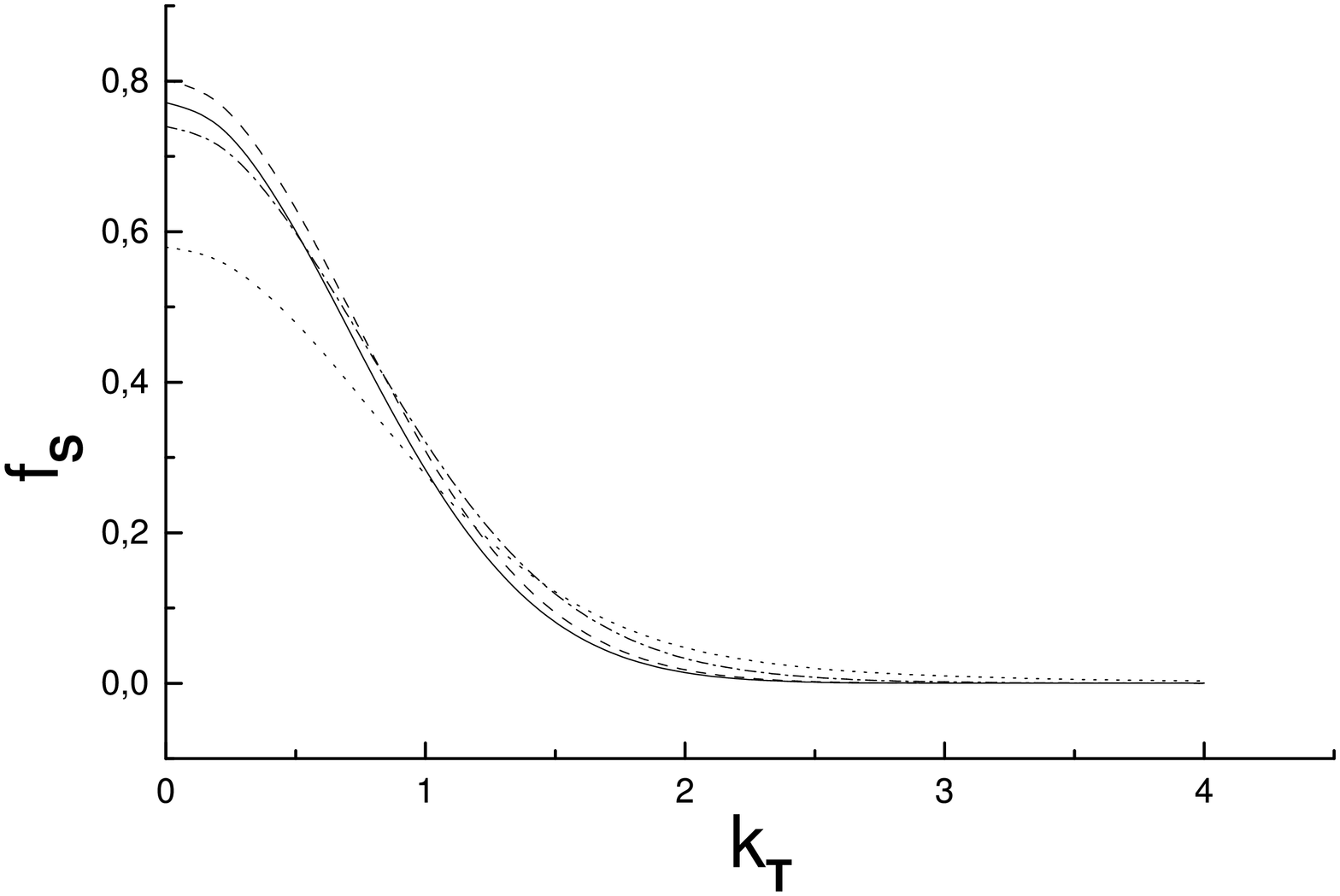}}
\subfigure{\includegraphics[angle=-90,width=0.43\textwidth]{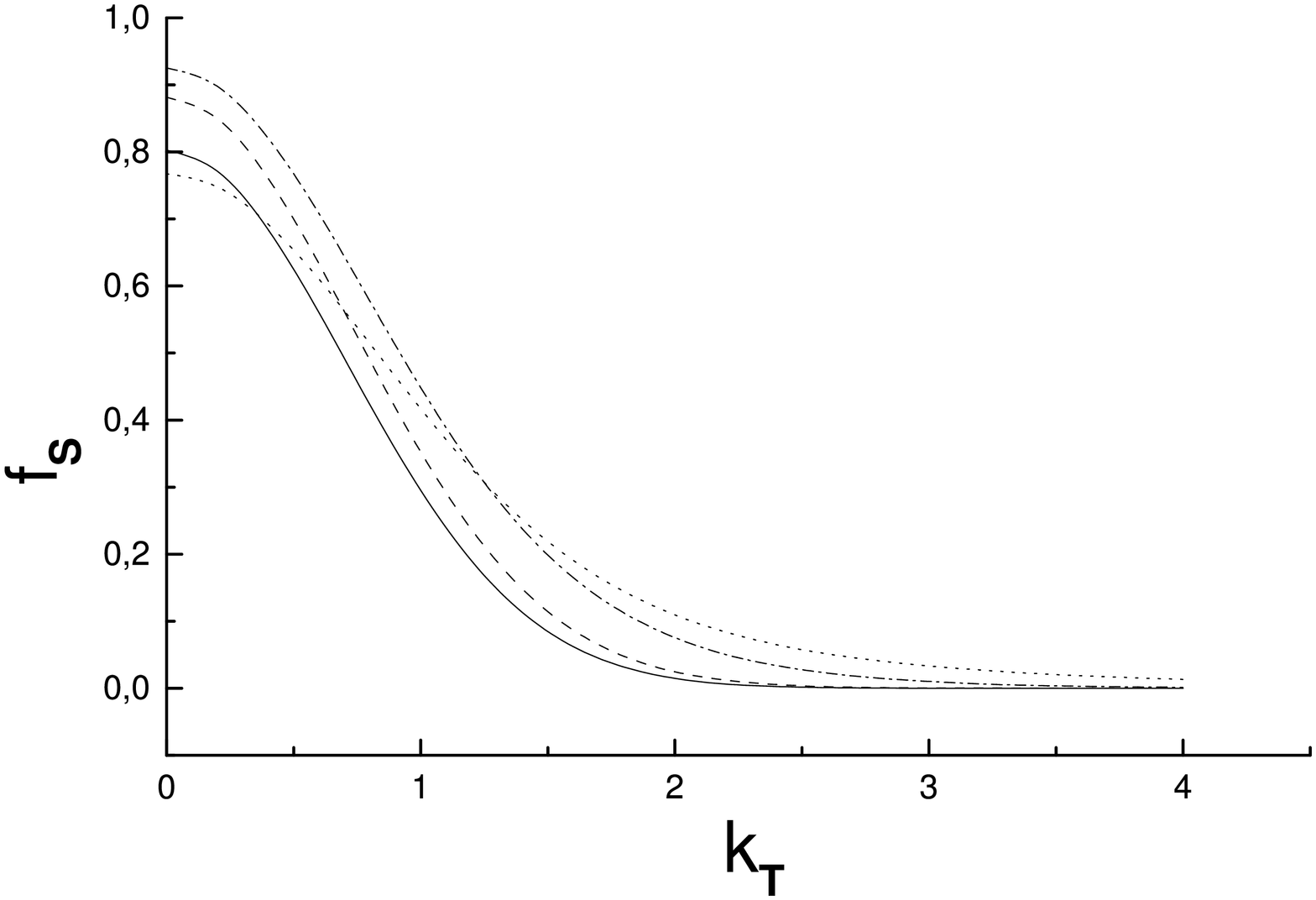}}\\
\subfigure{\includegraphics[angle=-90,width=0.43\textwidth]{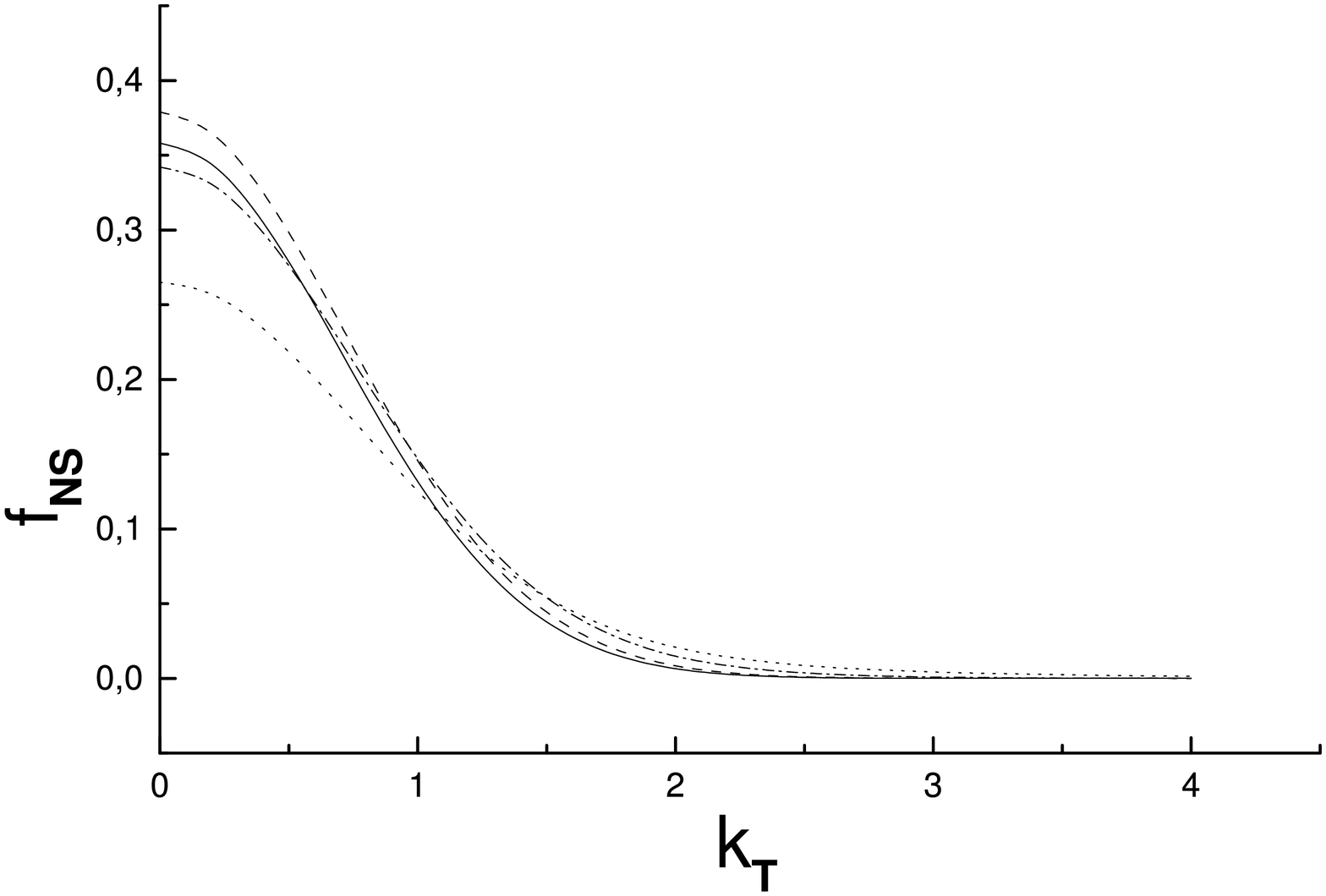}}
\subfigure{\includegraphics[angle=-90,width=0.43\textwidth]{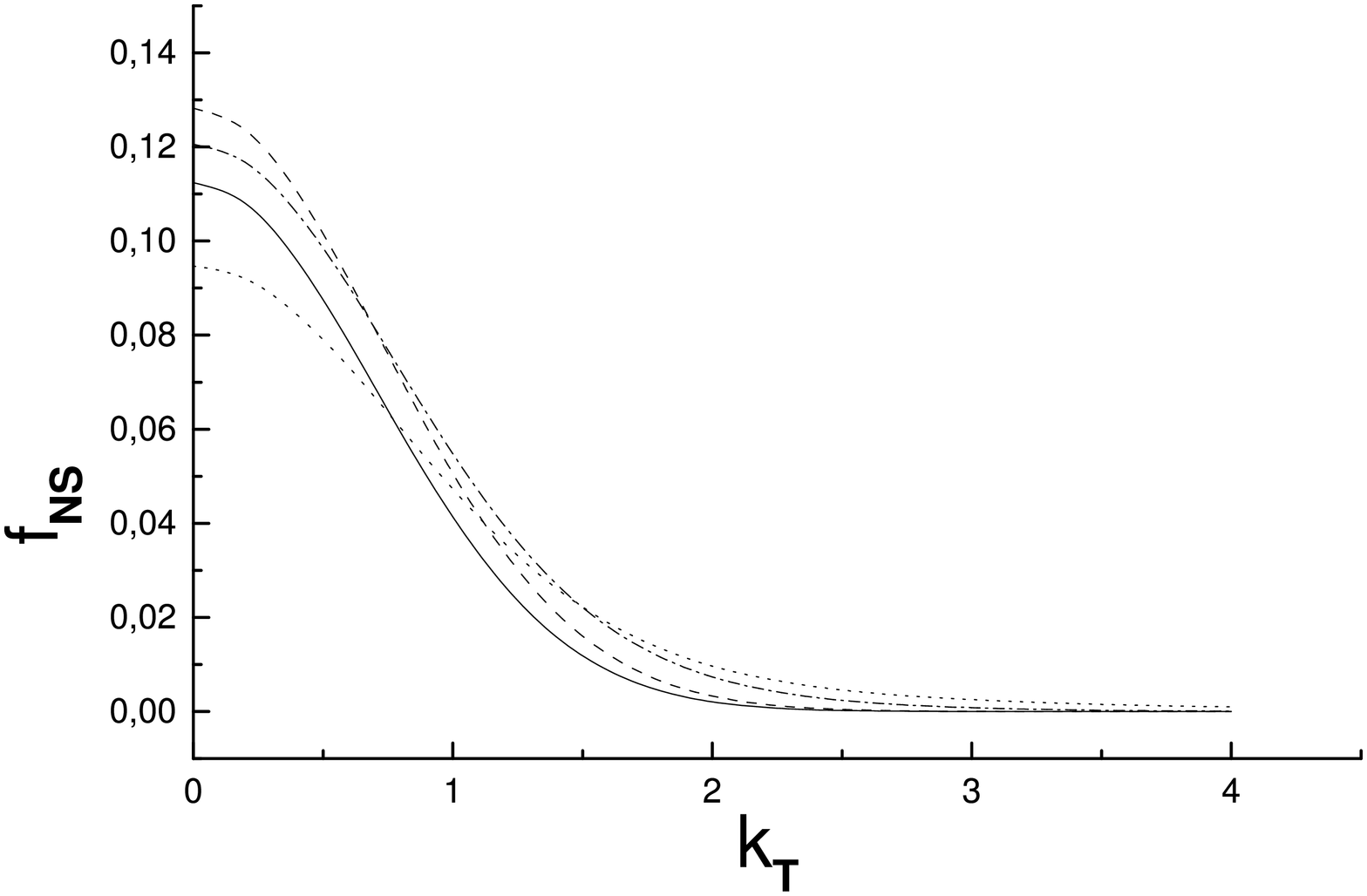}}
\end{center}
\caption{Transverse-momentum parton distributions in the pion following from the solution 
of the CCFM equation in the single loop approximation. The gluon, ${f}_G$, 
quark singlet, 
$f_S$, and quark non-singlet distributions, ${f}_{NS}$, are plotted as 
functions of the transverse momentum, $k_T$. The left and right parts are for $x=0.1$ and $x=0.01$, respectively.
The solid lines correspond to the initial scale of the GRS parameterisation, $Q^2=Q_0^2=0.26~{\rm GeV}^2$ 
\cite{GRS}, while the dashed, dash-dotted and dotted lines correspond to $Q^2=1~{\rm GeV}^2$, 
$10~{\rm GeV}^2$, and $100~{\rm GeV}^2$, respectively.  Parton distributions $f_i$ are 
in ${\rm GeV^{-2}}$ and transverse momentum $k_T$ is in ${\rm GeV}$.} \label{fig:ktdGRS}
\end{figure}

In Fig. \ref{fig:ktdGRS} we show the effect of the CCFM evolution on the $k_T$ 
distributions.  
We plot the unintegrated distributions $f_{G,S,NS}(x,k_T,Q)$ in 
the pion for $x=0.1$ and $x=0.01$, and for four values of the hard scale 
$Q^2$ varying from $Q_0^2$ to $Q^2=100 {\rm GeV}^2$. 
We have assumed here the Gaussian profile,
\begin{equation}
F(b)=\exp(-b^2 k_0^2/4),
\label{gauss}
\end{equation} 
which leads to Gaussian $k_T$ distributions at the the initial scale, $Q_0$,
proportional to $\exp(-k_T^2/k_0^2)$. The width parameter is taken to be
$k_0^2=1 {\rm GeV}^2$.  The broadening of the distributions with increasing magnitude 
of $Q^2$ and decreasing $x$ is  clearly visible.  It can also be seen that the effect is 
strongest for the gluons.  Important property of the broadening is significant modification 
of the Gaussian shape and development of the non-Gaussian long-range tail at large $k_T$. 

Strong increase of the average transverse momentum squared with increasing $Q^2$ does also indirectly 
affect $Q^2$ dependence of unintegrated distributions at $k_T=0$.   
In order to understand this dependence one may use the following approximate relation: 
\begin{equation}
f_i(x,k_T=0,Q) \sim {xp_i(x,Q^2)\over < k_T^2(x,Q^2)>_i}
\label{fkt0}
\end{equation}
where $p_i(x,Q^2)$ are the integrated distributions.  
(Equation (\ref{fkt0}) follows 
from 
approximating in  (\ref{fb2})   the distributions  
$\bar f_i(x,b,Q)$ by a Gaussian in $b^2$.)   One  expects that $f_i(x,k_T=0,Q)$ should decrease 
at large $Q^2$ due to strong increase of  $< k_T^2(x,Q^2)>_i$ with $Q^2$ that can be clearly seen from the 
from Figure (\ref{fig:ktdGRS}), particularly for the gluons.  In fact we  show in the next Section 
that $< k_T^2(x,Q^2)>_{i}$ should increase as $Q^2$ (modulo logarithmic effects). This 
increase is stronger than the potential increase of $xp_i(x,Q^2)$ caused by scaling 
violations.   
Initial increase of $f_G(x,k_T=0,Q)$  with increasing $Q^2$ for low values of $Q^2$,  
which can be clearly seen, particularly for low value of $x=0.01$ is caused by 
  the fact that at low $Q^2$ 
      $< k_T^2(x,Q^2)>_{evol,G}<<k_0^2$, i.e.  $< k_T^2(x,Q^2)>_{G}\sim k_0^2$ and so  
the $Q^2$ dependence of  $f_G(x,k_T=0,Q)$  is entirely driven by scaling violations 
of $xp_g(x,Q^2)=xg(x,Q^2)$.    

\begin{figure}[b]
\begin{center}
\includegraphics[width=13.5cm]{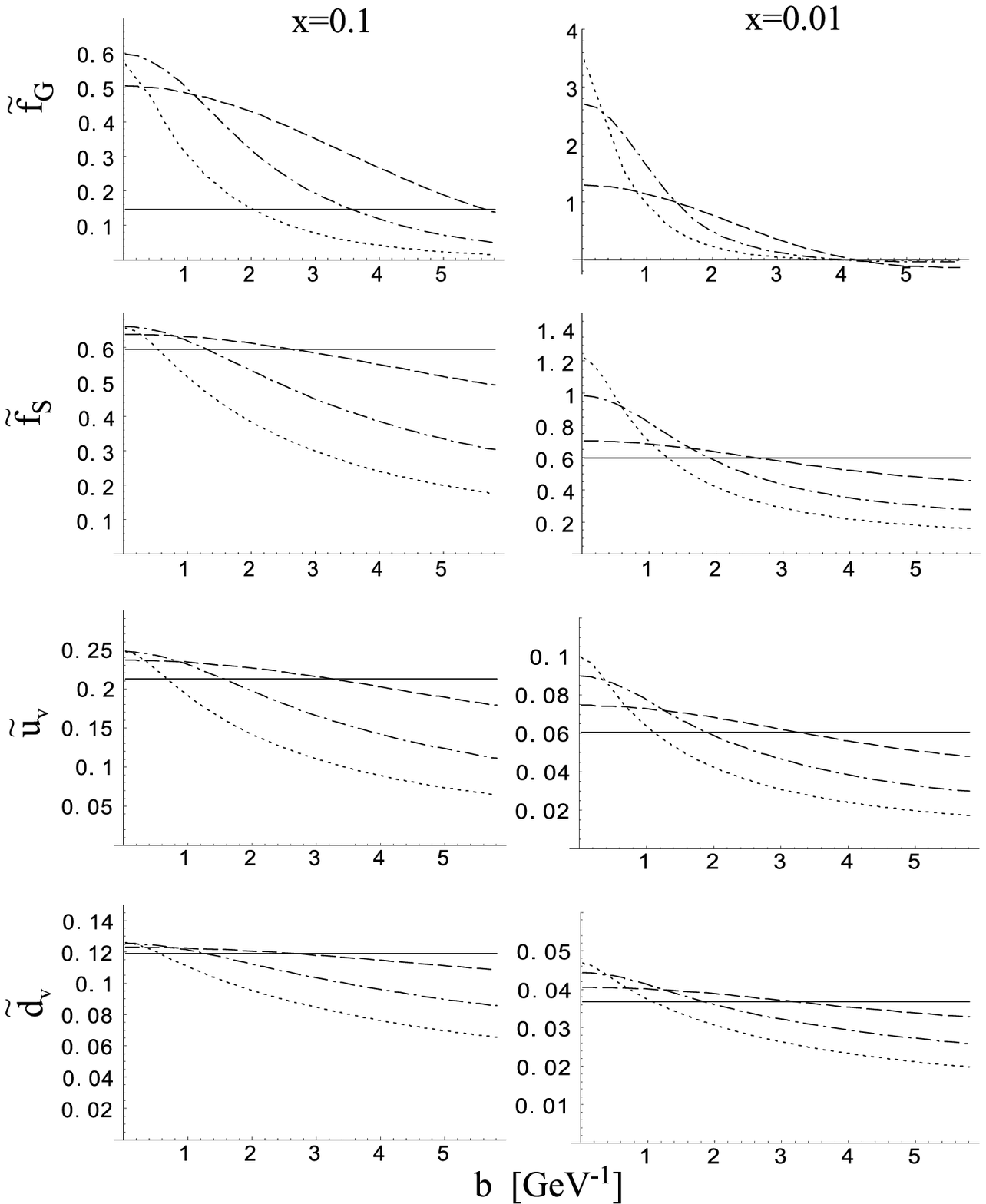}
\end{center}
\caption{Same as Fig. \ref{fig:bGRS} for the case of the nucleon, with the initial GRV 
parameterisation \cite{GRV}.} \label{fig:bGRV}
\end{figure}

\begin{figure}[b]
\begin{center}
\includegraphics[width=9cm]{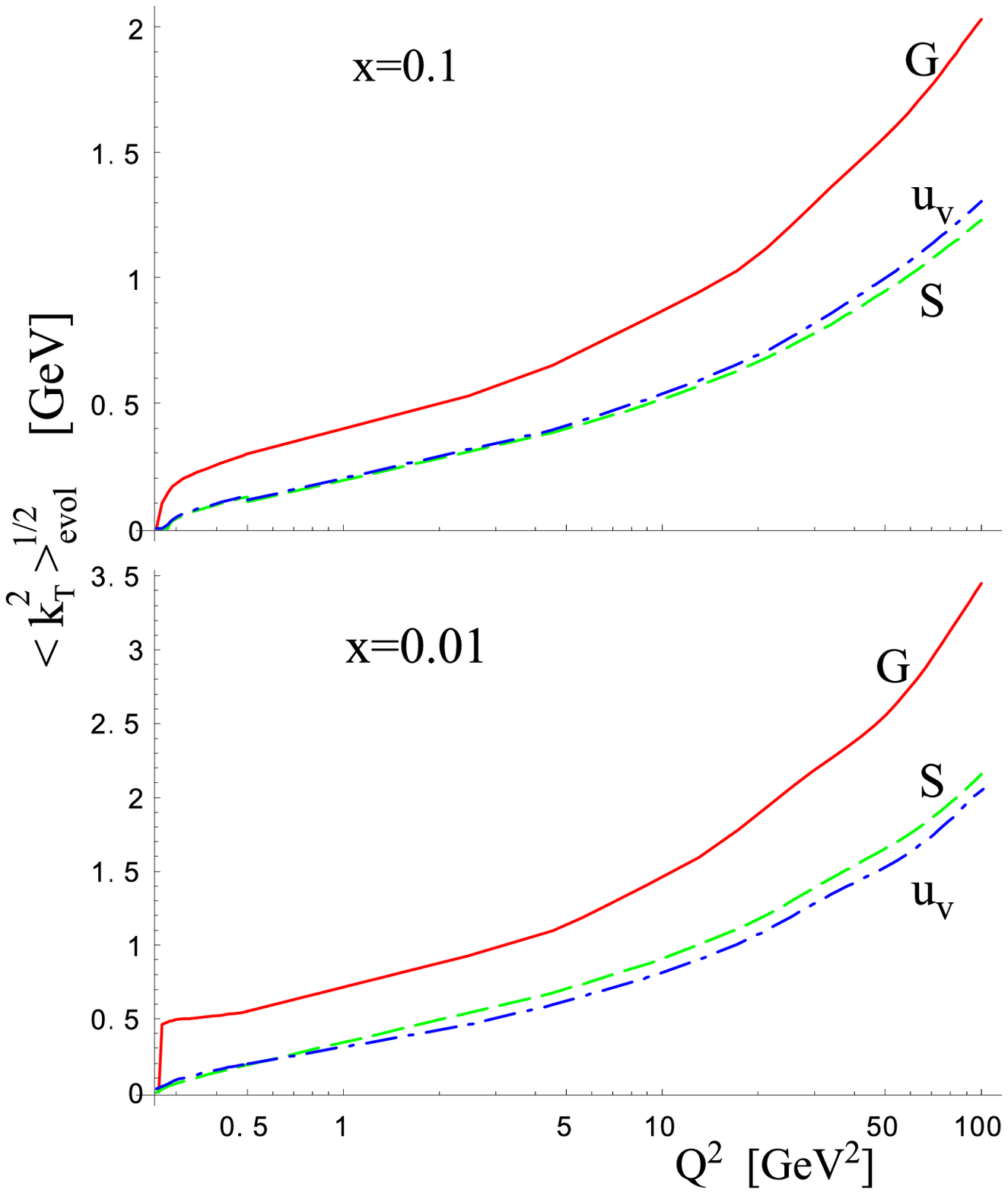}
\end{center}
\caption{Same as Fig. \ref{fig:bGRV} for the case of the nucleon, with the initial GRV 
parameterisation \cite{GRV}.}
\label{fig:ktGRV}
\end{figure}

\begin{figure}[b]
\begin{center}
\subfigure{\includegraphics[angle=-90,width=0.4\textwidth]{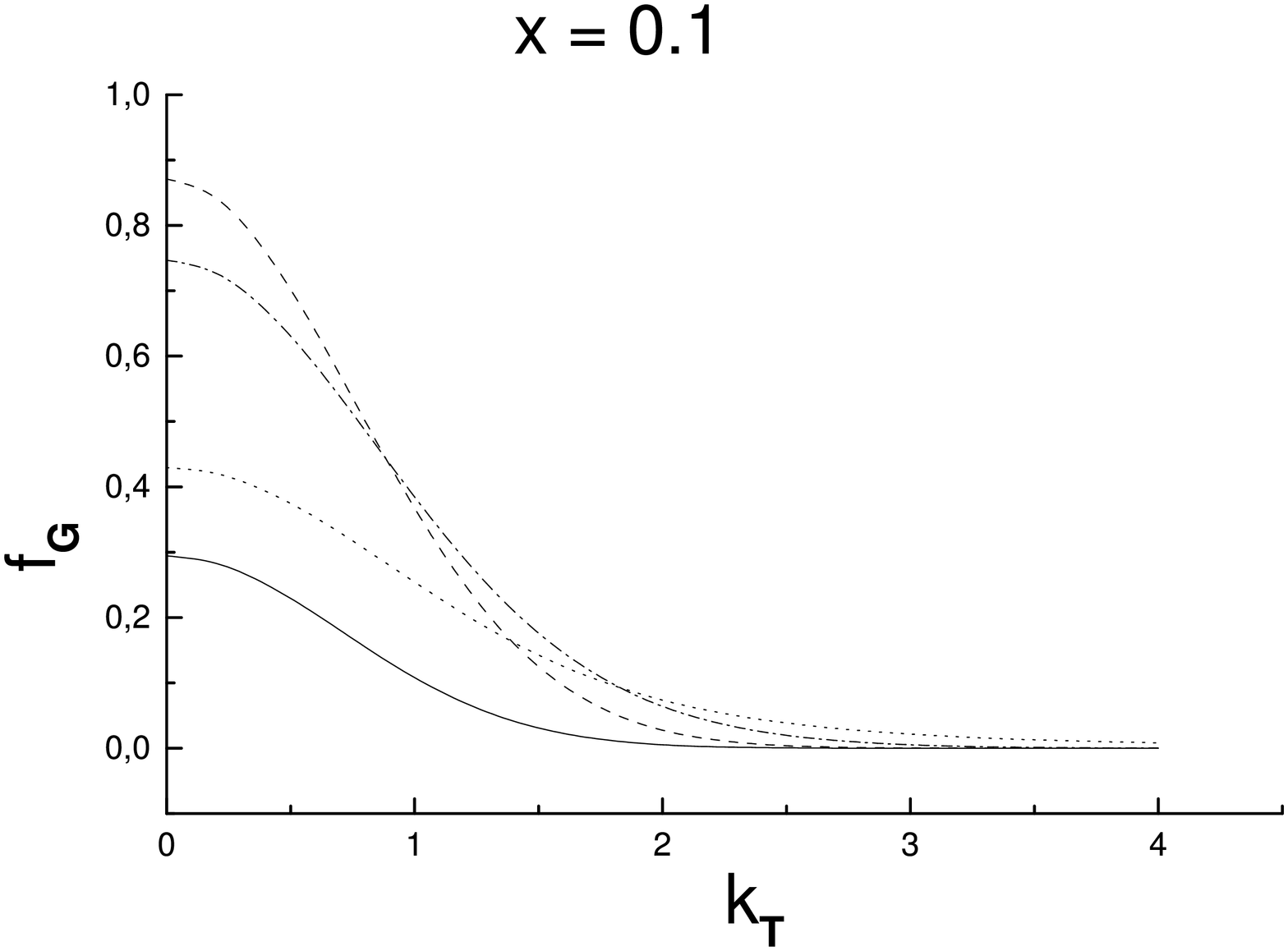}}
\subfigure{\includegraphics[angle=-90,width=0.4\textwidth]{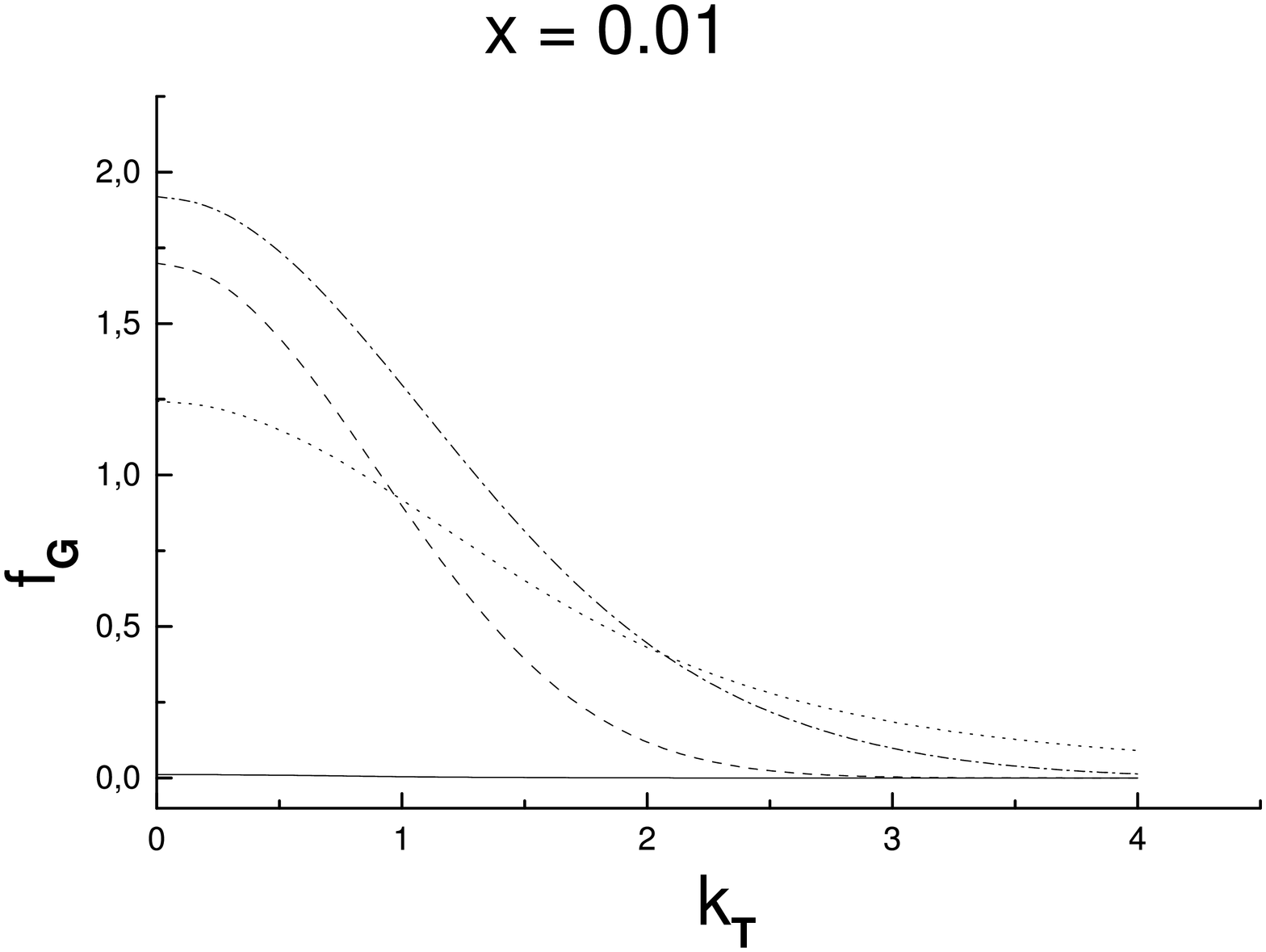}}\\
\subfigure{\includegraphics[angle=-90,width=0.4\textwidth]{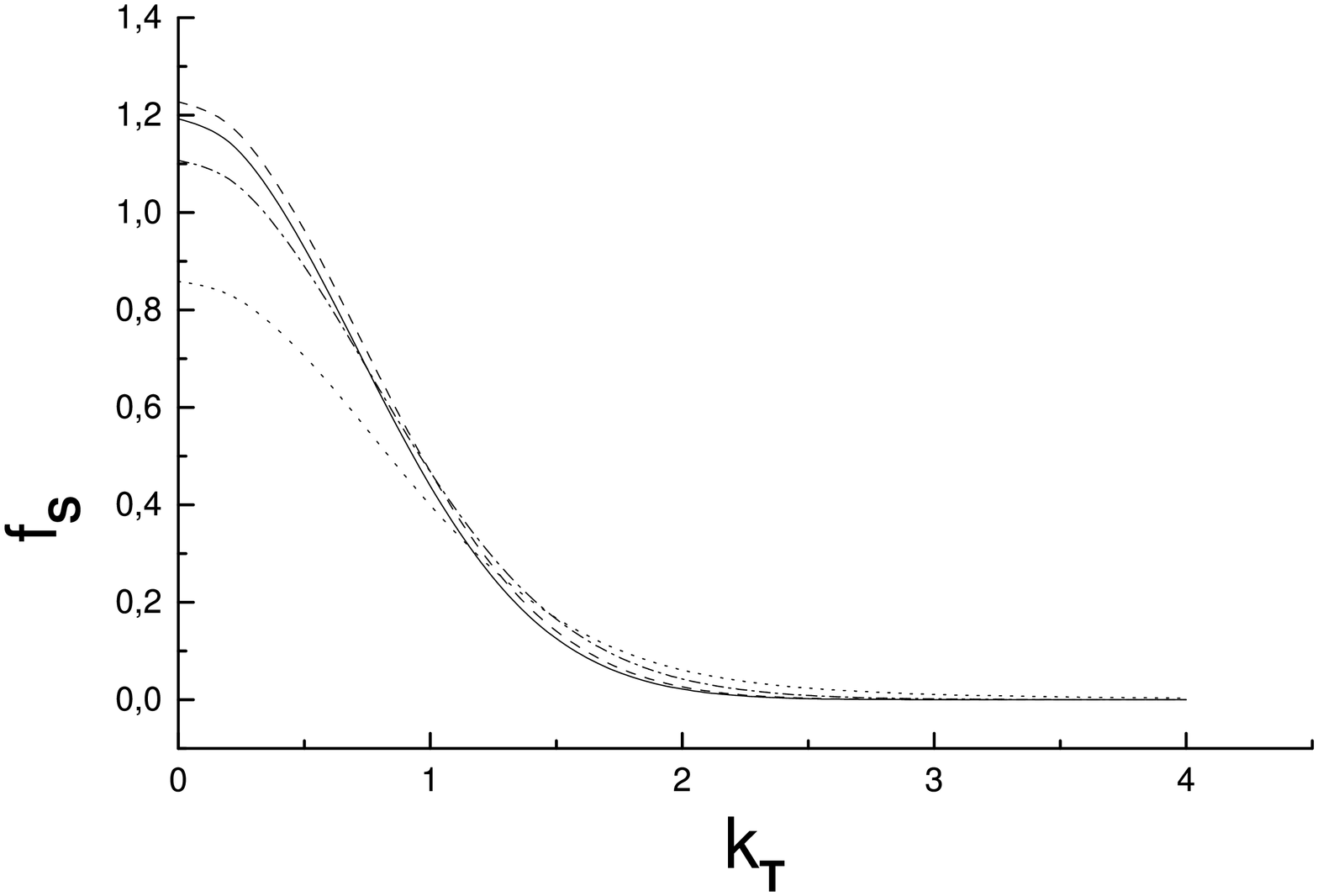}}
\subfigure{\includegraphics[angle=-90,width=0.4\textwidth]{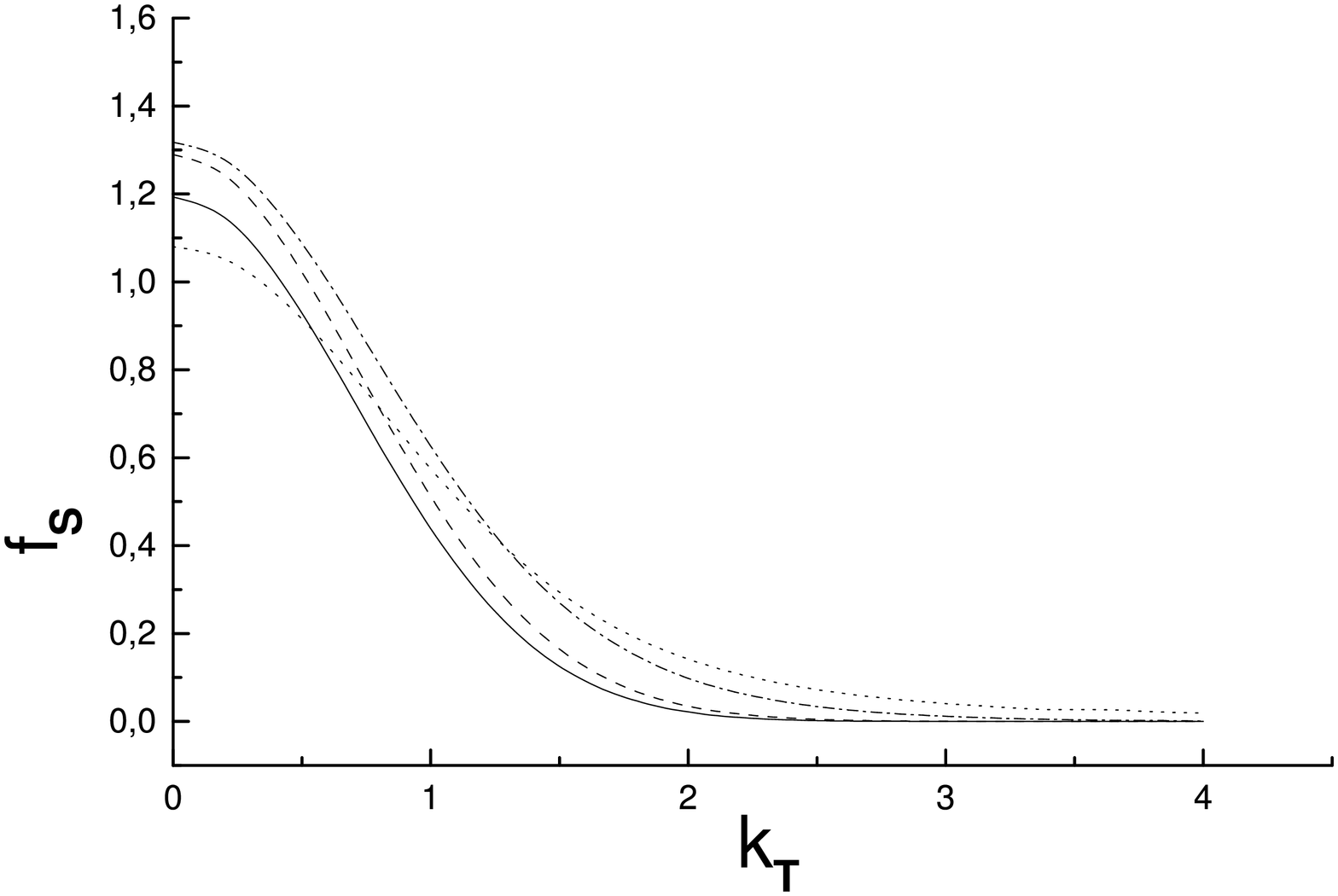}}\\
\subfigure{\includegraphics[angle=-90,width=0.4\textwidth]{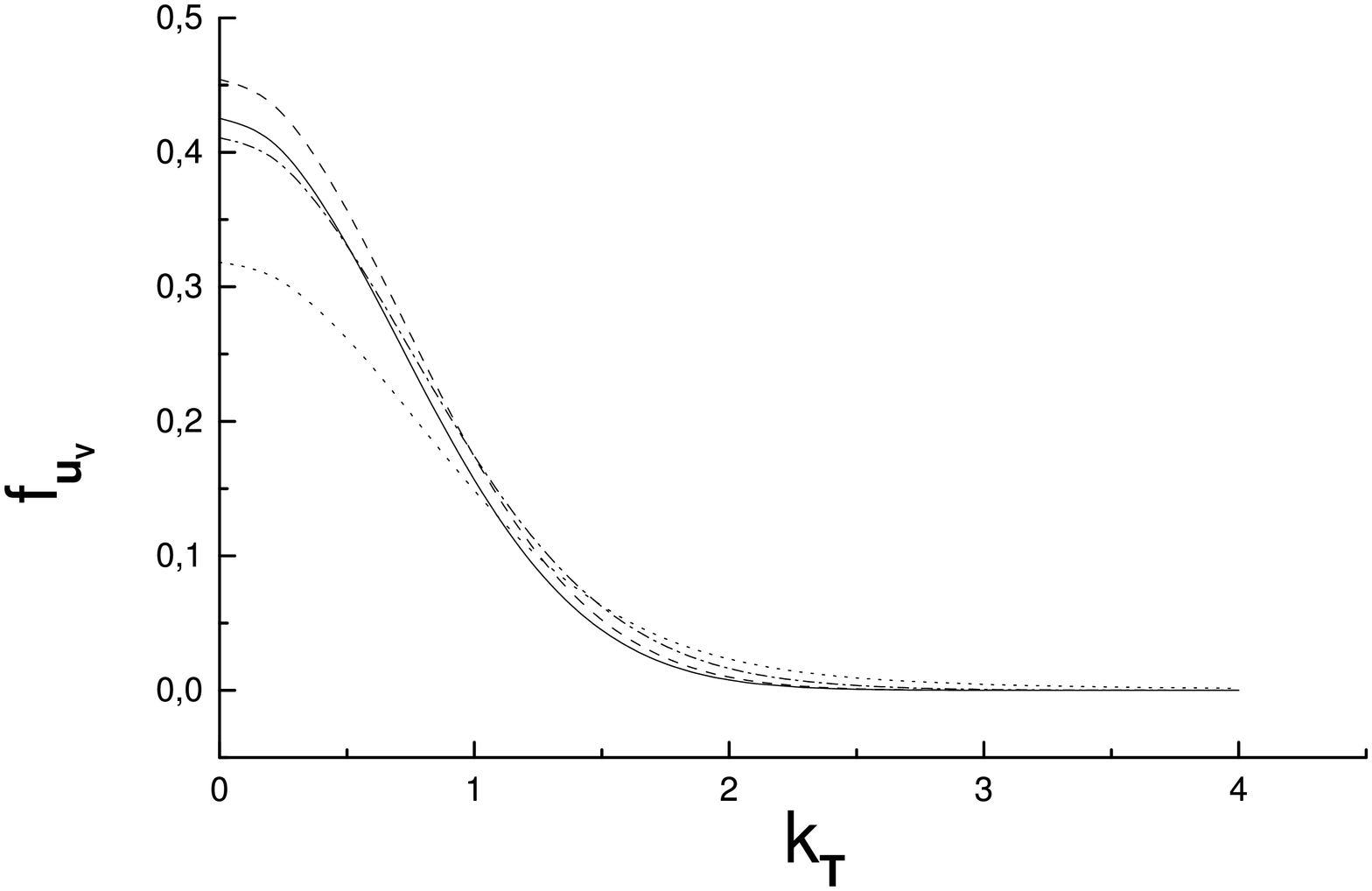}}
\subfigure{\includegraphics[angle=-90,width=0.4\textwidth]{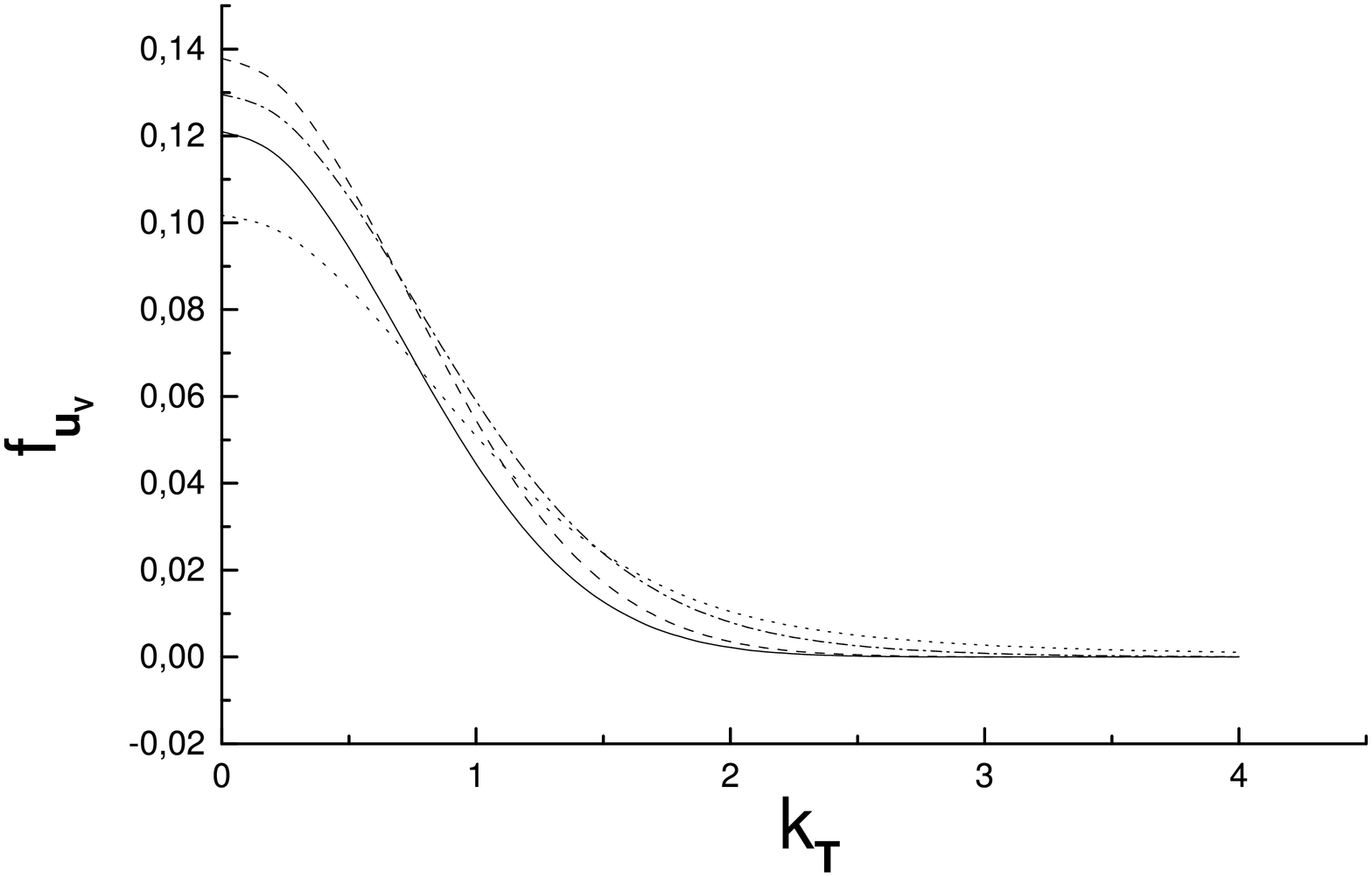}}
\end{center}
\caption{Same as Fig. \ref{fig:ktdGRS} for the case of the nucleon, with the initial GRV 
parameterisation \cite{GRV}.} \label{fig:ktdGRV}
\end{figure}

Figures \ref{fig:bGRV}-\ref{fig:ktdGRV} show the results analogous to Figs.~\ref{fig:bGRS}-\ref{fig:ktdGRS}
for the case of the nucleon, where the Gl\"uck-Reya-Vogt parameterisation of Ref.~\cite{GRV} is used at the initial
scale of $Q_0^2=0.26~{\rm GeV}^2$. The results are very similar to the case of the pion presented above.
For the case of the transverse-coordinate distributions we provide in Fig. \ref{fig:bGRV} the results for both 
the $u$ and $d$ valence quarks. For the nonsinglet distributions the other figures show the 
results for the $u$ valence quarks only, since the case 
of the $d$ valence quarks is very similar.


\section{Evolution of the $x$-moments}

The evolution equation for the $x$-moments of the non-singlet distribution, 
\begin{eqnarray}
\tilde{f}_{\rm NS}(n,b,Q^2)=\int_0^1 dx \,x^{n-1} \tilde{f}_{\rm NS}(x,b,Q), \label{momNS}
\end{eqnarray}
is particularly simple.  It immediately follows from Eq. (\ref{dccfmnsb}) that
\begin{eqnarray}
Q^2 \frac{\partial \tilde{f}_{\rm NS}(n,b,Q^2)}{\partial Q^2}=
\frac{\alpha_S(Q^2)}{2\pi} \int_0^1 dz P_{qq}(z) \left [ 
z^n J_0 \left ( (1-z) Qb) \right ) -1 \right ] \, \tilde{f}_{\rm NS}(n,b,Q), \nonumber \\ \label{momNSevol}
\end{eqnarray}
which can be transformed into 
\begin{eqnarray}
Q^2 \frac{\partial \log \left ( \tilde{f}_{\rm NS}(n,b,Q^2) \right ) }{\partial Q^2}=
\frac{\alpha_S(Q^2)}{2\pi} \int_0^1 dz \frac{4}{3} \frac{z^2+1}{1-z} \left [ 
z^n J_0 \left ( (1-z) Qb) \right ) -1 \right ] . \label{momNSevol2}
\end{eqnarray}
At $b=0$ this equation reproduces the standard evolution of non-singlet moments for the DGLAP equations.
Next, we differentiate with respect to $b^2$ at $b=0$ on both sides of Eq. (\ref{momNSevol2}) and evaluate
the $z$ integral on the right-hand side. The simple result is  
\begin{eqnarray}
\frac{\partial \langle k^2_T(n,Q^2) \rangle_{\rm NS, evol} }{\partial Q^2}=
\frac{\alpha_S(Q^2)}{2\pi} c_n, \label{momNSevol3}
\end{eqnarray}
with the definition
\begin{eqnarray}
\langle k^2_T(n,Q^2) \rangle_{\rm NS, evol}=
\frac{\int d^2{\bf k}_T \, k_T^2 \tilde{f}_{\rm NS}(n,k_T,Q)}
{\int d^2{\bf k}_T \, \tilde{f}_{\rm NS}(n,k_T,Q}
\label{ktdef}
\end{eqnarray}
and 
\begin{eqnarray}
c_n=\frac{4}{3}\left ( \frac{1}{n+1}-\frac{1}{n+2} + \frac{1}{n+3}-\frac{1}{n+4} \right ) .
\label{cndef}
\end{eqnarray}
The integration of Eq. (\ref{momNSevol3}) with the condition $\langle k^2_T(n,Q_0^2)
 \rangle_{\rm NS, evol}(Q_0^2)=0$
yields the final result
\begin{equation}
\langle k^2_T (n,Q^2) \rangle_{\rm NS, evol}=
c_n \int_{Q^2_0}^{Q^2}dq^2 {\alpha_s(q^2)\over 2 \pi},
\label{momfin}
\end{equation}
where 
\begin{equation}
\int_{Q^2_0}^{Q^2}dq^2 {\alpha_s(q^2)\over 2 \pi}=
\frac{2 \Lambda_{QCD}^2}{\beta_0}  \left [ {\rm li}\left ( \frac{Q^2}{\Lambda_{QCD}^2} \right ) - 
{\rm li}\left ( \frac{Q_0^2}{\Lambda_{QCD}^2} \right ) , \right ] 
\label{momfine}
\end{equation}
with $\beta_0=11N_c/9-2N_f/3$ and $ {\rm li}(z)=\int_0^z dt/\log t$. Obviously, 
the moments for all values of $n$ are growing with $Q^2$, which once again is 
a manifestation
of the advocated spreading in $k_T$. 

The analysis for the gluon and singlet quarks is more complicated 
due to mixing and is discussed in the Appendix. The evolution parts $<k^2_T(n,Q^2)>_{\rm i,evol}$, 
with $i={\rm G, S}$, 
are equal to 
\begin{equation}
<k^2_T(n,Q^2)>_{\rm i ,evol}=-4{f_{\rm i ,evol}^{\prime}(n,Q^2)\over f_{\rm i, evol}(n,Q^2)} ,
\label{ktavsg}
\end{equation}
with the functions $f_{i,evol}^{\prime}(n,Q^2)$ and $f_{i, evol}(n,Q^2)$ defined in 
Eqs.~(\ref{spsg}) and (\ref{fsgn}) in the Appendix.  From those formulas and from 
equation (\ref{momfin}) it follows that $<k^2_T(n,Q^2)>_{\rm i}$ increases as $Q^2$, 
modulo logarithmic effects.  This increase is caused by the fact that 
$-f_{\rm i ,evol}^{\prime}(n,Q^2)$ is found to increase as $Q^2$ modulo logarithmic effects 
caused by QCD scaling violations.  Similar strong increase should also hold for 
$-d/db^2\tilde f_i(b,x,Q)|_{b=0}$ controlling $Q^2$ dependence of $<k_T^2(x,Q^2)>_{i,evol}$

\begin{figure}[b]
\begin{center}
\includegraphics[width=9cm]{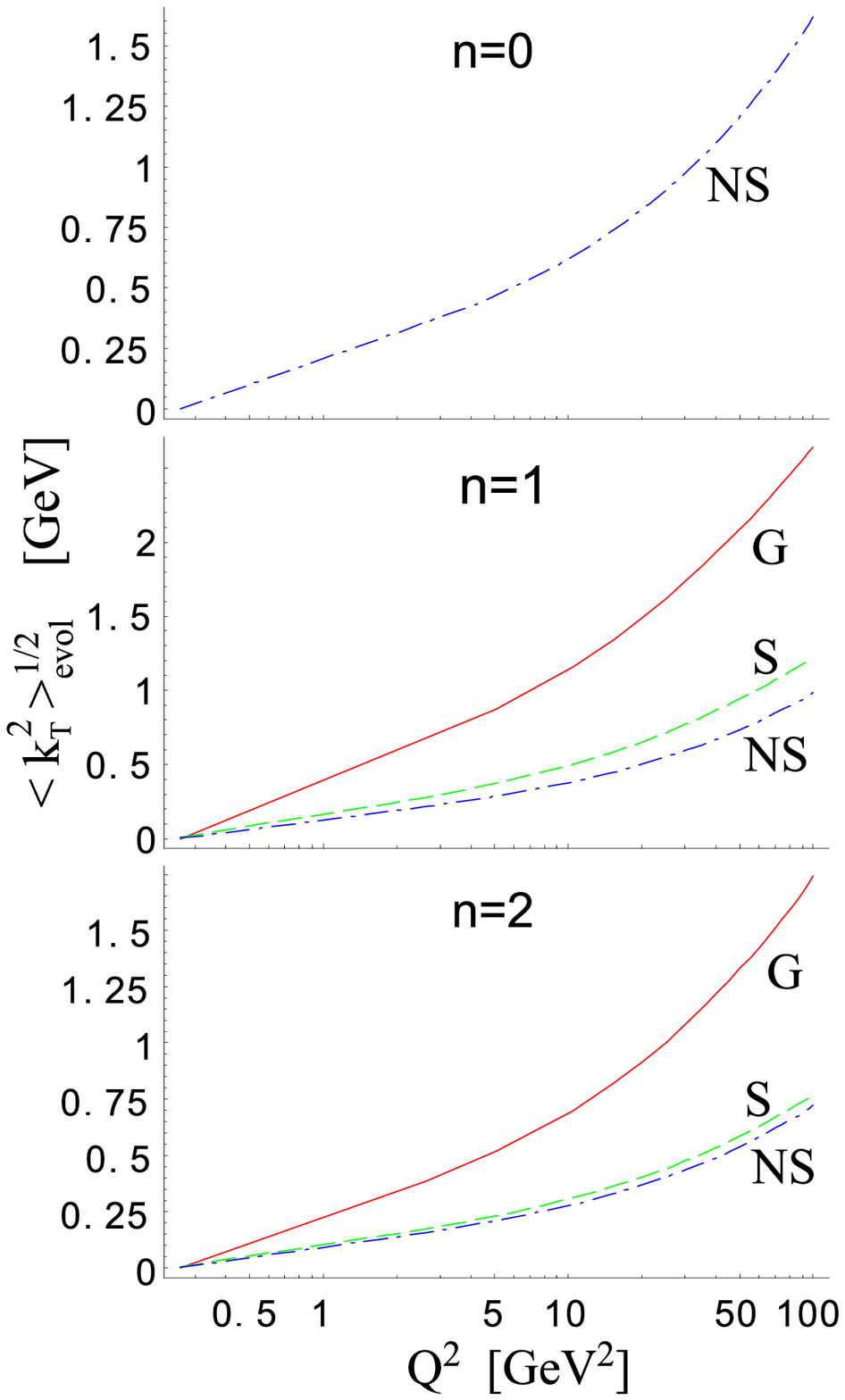}
\end{center}
\caption{The dynamically-generated root mean squared transverse momentum of unintegrated partonic distribution 
functions in the pion corresponding to the $x$ moments of the distributions, 
plotted as a function of $Q^2$ for the moment index $n=0$ (top), $n=1$ (middle), and $n=2$ (bottom). 
The solid, dashed, and dot-dashed lines denote gluons, singlet quarks, and 
non-singlet quarks. For $n=0$ only the non-singlet distribution makes sense. The CCFM equations in the single-loop 
approximation with the initial GRS condition \cite{GRS} at the scale 
$Q^2=Q_0^2=0.26~{\rm GeV}^2$ are used for the evolution. The analogous plot for the nucleon with the GRV 
parameterisation gives identical results for the nonsinglet distributions, and practically identical
for the gluons and singlet quarks. }
\label{fig:ktmGRS}
\end{figure}

The results for the dynamically-generated root mean squared transverse momentum for the $x$ moments of the distributions 
in the pion, $<k^2_T(n,Q^2)>_{\rm i ,evol}$, are shown in Fig.~\ref{fig:ktmGRS}. For $n=0$ only the nonsinglet distribution 
is shown, since the gluon and singlet equations involve singularities and do not make sense in this case.  
The solid, dashed, and dot-dashed lines denote gluons, singlet quarks, and 
non-singlet quarks. Again we use the initial GRS condition \cite{GRS} at the scale 
$Q^2=Q_0^2=0.26~{\rm GeV}^2$. The analogous plot for the nucleon with the GRV 
parameterisation \cite{GRV} gives identical results for the nonsinglet distributions, which follows immediately from 
Eq.~(\ref{ktdef}) not involving the initial values of the moments. 
For the gluons and singlet quarks the results for the pion and the nucleon, although not strictly
equal (see the Appendix), are practically identical. 
The results displayed in Fig.~\ref{fig:ktmGRS} once again show the main observation of this paper, namely, a 
large spreading of the distributions in process of the single-loop CCFM evolution.

\section{Parton-parton luminosity function}

\begin{figure}[b]
\begin{center}
\includegraphics[angle=-90,width=13.5cm]{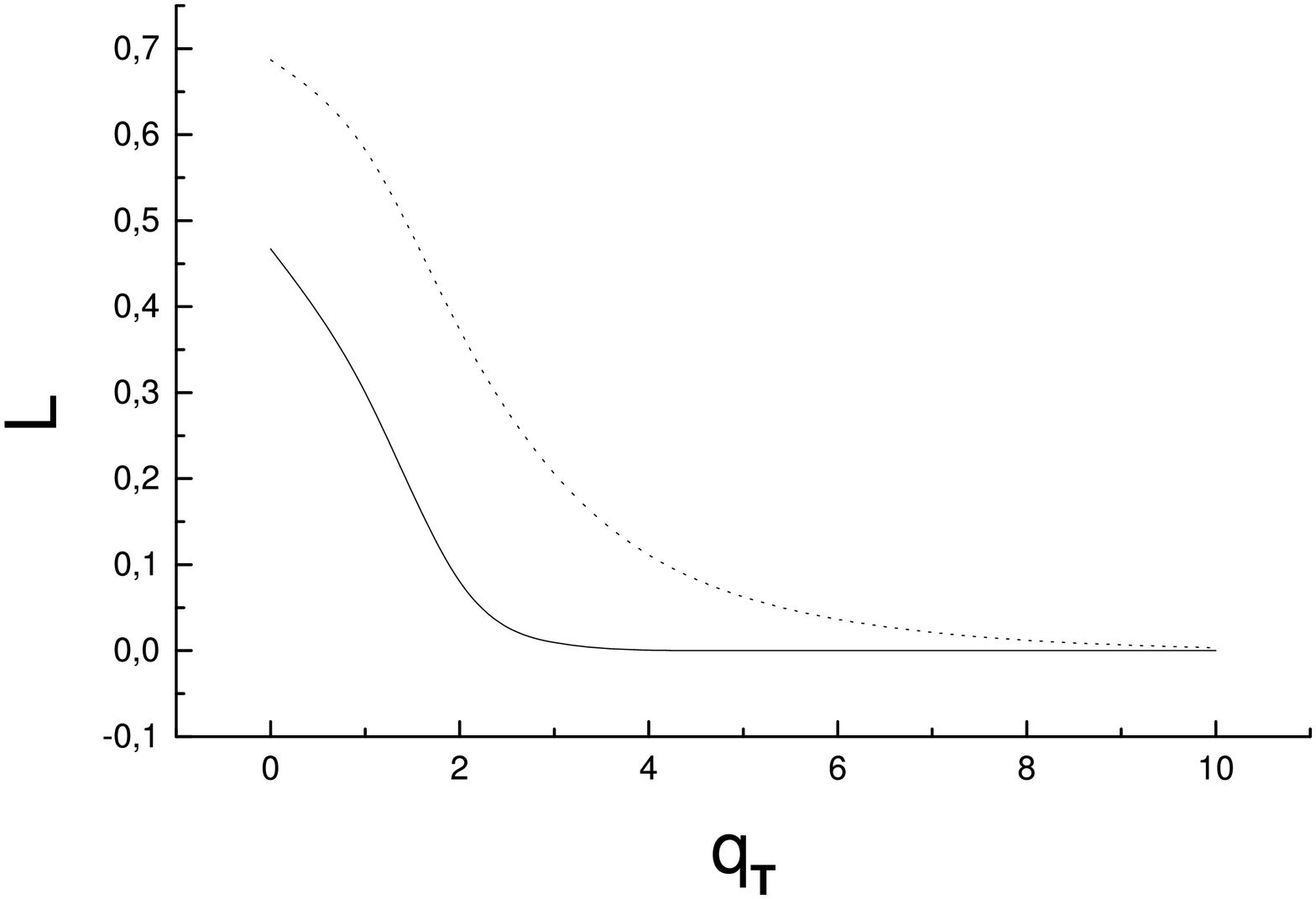}
\end{center}
\caption{The parton luminosity function  corresponding to the valence $u$ quark - gluon 
collisions plotted as the function of the transverse momentum 
$q_{T}$ equal to the vector sum of the transverse momenta of the colliding partons.  
The value of the hard scale 
$Q$ is set to the transverse momentum $p_{\gamma T}$ of the produced 
photon.  The two curves correspond 
to two values of the total $\bar p p$ CM energy $\sqrt{s}$ and different values of $p_{\gamma T}$:
$\sqrt{s} = 24.3 {\rm GeV}$, $p_{\gamma T} = 4 {\rm GeV}$ (dashed line) and 
$\sqrt {s} = 1800 {\rm GeV}$, $p_{\gamma T} = 10 {\rm GeV}$ (full line).  Results for 
$\sqrt{s} = 24.3 {\rm GeV}$ were multiplied by $10$.  The parton luminosity function $L$ is 
in ${\rm GeV^{-2}}$ and transverse momentum $q_T$ is in ${\rm GeV}$.}
\label{fig:lumino}
\end{figure}

The knowledge of unintegrated distributions in the $b$ representation is useful for calculations of the parton-parton 
luminosity function corresponding to the collision of parton $a$ with the longitudinal momentum fraction $x_{1}$ and 
the transverse momentum fraction $k_{1T}$ with parton $b$ with the longitudinal momentum fraction $x_{2}$ and 
the transverse momentum fraction $k_{2T}$. The luminosity function is defined as \cite{KMR1}: 
\begin{equation}
 L_{ab}(x_{1}, x_{2}, q_{T}, Q)=\int \frac {d^{2}{\bf k_{1T}}d^2{\bf k_{2T}}}{\pi}
  f_{a}(x_{1}, k_{1T}, Q) f_{b}(x_{2}, k_{2T}, Q) \delta ^{(2)}({\bf k_{1T}}+ {\bf k_{2T}} - {\bf q_{T}}).
\label{lumi1}
\end{equation}
Substituting in Eq.~(\ref{lumi1})  the following representation of the 
$\delta$ function,
\begin{equation}
\delta ^{(2)}({\bf k_{1T}}+ {\bf k_{2T}} - {\bf q_{T}})={1\over (2 \pi)^2} 
\int d^2{\bf b} \exp[({\bf k_{1T}}+ {\bf k_{2T}} - {\bf q_{T}}){\bf b}],
\label{delta2}
\end{equation} 
we find that the integral defining the parton-parton luminosity in Eqn.~(\ref{lumi1}) 
can be directly expressed in terms of the 
unintegrated distributions in the $b$ representation, 
\begin{equation}
L_{ab}(x_{1}, x_{2}, q_{T}, Q)= 
\int_0^{\infty}db^2 f_{a}(x_{1}, b, Q) f_{b}(x_{2}, b, Q) J_0(bq_{T}).
\label{lumi2}
\end{equation}
In Fig. \ref{fig:lumino} we show the luminosity function corresponding to the valence $u$ quark-gluon collisions. 
This quantity is relevant for the description of prompt photon production
for which the dominant subprocess is $u + g \rightarrow u + \gamma $. We set 
$Q = p_{\gamma T}$ and 
$x_{1} = x_{2} = {2 p_{\gamma T}}/{\sqrt {s}}$ where $\sqrt {s}$ is the CM energy of the $p \bar p$ 
collisions. The two curves correspond  
to $\sqrt{s} = 24.3 {\rm GeV}$, $p_{\gamma T} = 4 {\rm GeV}$ and 
$\sqrt{s} =  1800 {\rm GeV}$, $p_{\gamma T} = 10 {\rm GeV}$.  We can see that the luminosity 
function is increasing with the energy, which simply  reflects the increase of  
the unintegrated gluon  distributions with decreasing 
$x={2 p_{\gamma T}}/{\sqrt {s}}$. 

\section{Summary and conclusions.}
In this paper we have performed an analysis of the unintegrated parton distributions 
$f_i(x,k_T,Q)$ of the pion and 
nucleon using the CCFM equation in the single loop approximation.  For the integrated
distributions this approximation is equivalent 
to the LO DGLAP evolution, and thus should be adequate in the region of large and moderately small values of 
$x$ ({\em i.e.} $x> 0.01$ or so).  
Important  merit of the  CCFM framework  is the fact that once the input $x$ and $k_T$ 
parton distributions are provided at the reference scale $Q_0^2$, the 
CCFM evolution 
generates the $x$ and $k_T$ distributions for arbitrary $Q^2>Q_0^2$ and arbitrary values of 
$k_T$, including the region of low $k_T$  down to the point $k_T=0$.  This framework 
 does therefore 
provide a useful extension of the approximate DGLAP - based treatment of the unintegrated 
distributions discussed in \cite{DDT,KMR1}.     
We have extended the original CCFM equation 
by including the quarks and non-singular parts of the splitting functions and 
used the transverse-coordinate representation by introducing 
the parton distribution $\bar f_i(x,b,Q)$, related to the unintegrated 
parton distributions $f_i(x,k_T,Q)$  through the 
Fourier-Bessel transform.   The usefulness of this representation is related 
to the fact that the transverse coordinate $b$ is conserved through the CCFM evolution 
in the single loop approximation.  The average transverse momenta squared of the partons are determined 
by $d \bar f_i(x,b,Q)/db^2|_{b=0}$.      
 We have studied the impact of the QCD evolution on the shape of the 
$b$ profiles and on the broadening of the $k_T$ distributions. We have quantified 
increase of the average transverse momentum with increasing magnitude of the 
hard scale $Q$ and/or with decreasing $x$ and we gave semianalytic insight into this
increase.  The average transverse momenta squared of the partons were found to 
increase as $Q^2$ with increasing $Q^2$, modulo logarithmic effects related to the QCD 
evolution.   We have also pointed out that the parton distributions in $b$ representation 
determine the parton luminosity functions for fixed 
sum of the transverse momenta of the partons.  The unintegrated distributions $f_i(x,k_T,Q)$ 
(or $f_i(x,b,Q)$) obtained in our paper can be used for the theoretical 
description of the processes 
which are sensitive to the transverse momenta of the partons.

\section*{Acknowledgements}
This research was partially supported
by the Polish Committee for Scientific Research (KBN) grants no. 2P03B 05119 and 5P03B 14420.

\appendix    
\section*{Appendix}

The average value of the transverse momentum  $<k_T^2(n)>_i$ is given by
\begin{equation}
 <k_T^2(n,Q^2)>_i=-4{f_i^{\prime}(n,Q^2)\over f_i(n,Q^2)}
\label{ptn1}
\end{equation}
where 
\begin{eqnarray}
f_i^{\prime}(n,Q^2)&=&{d f_i(n,b^2,Q^2)\over db^2}|_{b^2=0}, \nonumber \\
f_i(n,Q^2)&=&f_i(n,b^2=0,Q^2),
\label{deriv}
\end{eqnarray}
and $f_i(n,b^2,Q^2)$ are the moment functions,
\begin{equation}
f_i(n,b^2,Q^2)=\int_0^1 dx x^{n-1}\bar f_i(x,b,Q).
\label{mfdef}
\end{equation}
The index $i$ corresponds to the gluon, quark singlet, or quark non-singlet 
distributions.
In the case of the quark NS distributions we just have
\begin{eqnarray}
<k_T^2(n,Q^2)>_{\rm NS}&=& <k_T^2(n,Q_0^2)>_{\rm NS}+\int_0^1dzz^{n}(1-z)^2P_{qq}(z)
\int_{Q_0^2}^{Q^2}dq^2 {\alpha_s(q^2)\over 2\pi}
\nonumber \\ 
&=&<k_T^2(n,Q_0^2)>_{\rm NS}+<k_T^2(n,Q_0^2)>_{\rm NS,evol}
\label{ptnns2}
\end{eqnarray}
In what follows we derive analytic expression for the dynamical component 
 $<k_T^2(n,Q^2)>_{i, {\rm evol}}$ 
of  $<k_T^2(n,Q^2)>_i$ corresponding to the gluon and quark singlet distributions. 
To this aim we start from the system of the evolution equations: 
\begin{equation}   
 Q^2{d\hat f^{\prime}(n,Q^2)\over dQ^2} = {\alpha_s(Q^2)\over 2\pi} 
\left[\hat P(n) \hat f^{\prime}(n,Q^2) -{Q^2\over 4}\hat P^{\prime}(n)\hat f(n,Q^2)\right],
\label{evold}
\end{equation}
obtained from Eqs.~(...) by differentiation with respect to $b^2$. 
The components of the vector $ \hat f^{\prime}(n,Q^2)$ are $ f_{S}^{\prime}(n,Q^2)$ 
and  $f_{G}^{\prime}(n,Q^2)$, while the components of the matrices $\hat P(n)$ are equal to the 
moments of the splitting functions defining DGLAP  evolution of the moments of 
singlet and gluon distributions, namely,
\begin{eqnarray}
\hat P_{qq}(n)&=&\int_0^1 dz P_{qq}(z)[z^{n}-1], \nonumber \\
\hat P_{gg}(n)&=&\int_0^1 dz \{zP_{gg}(z)[z^{n-1}-1]-zP_{qg}\}, \nonumber \\
\hat P_{qg,gq}(n)&=&\int_0^1 dz P_{qg,gq}(z) z^n.
\label{hatpn}
\end{eqnarray}
The explicit expressions for 
the functions $\hat P_{ij}(n)$ are 
\begin{eqnarray}
\hat P_{qq}(n)&=&{4\over 3}\left[{3\over 2}-{1\over  n+1} - {1\over  n+2} -2S(n)\right], \nonumber \\
\hat P_{qg}(n)&=&N_f\left[{1\over  n+1}-{2\over n+2} + {2\over n+3}\right], \nonumber \\
\hat P_{gq}(n)&=& {4\over 3}\left[{2\over n}-{2\over n+1}+{1\over  n+2}\right], \nonumber \\
\hat P_{gq}(n)&=&
6\left[{11\over 12}-S(n)+ {1\over n}-{2\over n+1}+{1\over  n+2} - {1\over  n+3} \right]-
{N_f\over 3},
\label{hpijn}
\end{eqnarray}
where 
\begin{equation}
S(n)=\int_0^1dz{1-z^n\over 1-z}=\sum_{k=1}^n \frac{1}{k}.
\label{sn}
\end{equation}
Similarly, the components of the matrices $\hat P^{\prime}(n)$
are defined as
\begin{equation}
\hat P_{ab}^{\prime}(n)=\int_0^1 dz (1-z)^2P_{ab}(z) z^n, 
\label{hatppn}
\end{equation}
with the explicit expressions given by
\begin{eqnarray}
\hat P_{qq}^{\prime}(n)&=& {4\over 3}\left[{1\over  n+1}-{1\over  n+2}+{1\over  n+3}-
{1\over  n+4}\right] , \nonumber \\
\hat P_{qg}^{\prime}(n)&=& N_f\left[{1\over  n+1}-{4\over n+2}+{7\over n+3}-
{6\over n+4} +{2\over n+5}\right], \nonumber \\
\hat P_{gq}^{\prime}(n)&=&
{4\over 3}\left[{2\over n}-{6\over n+1}+{7\over n+2}-
{4\over n+3} +{1\over n+4}\right], \nonumber \\
\hat P_{gg}^{\prime}(n)&=&6\left[{1\over n} -{3\over n+1}+{5\over n+2}-
{5\over n+3} +{3\over n+4}-{1\over  n+5}\right].
\label{hppijn}
\end{eqnarray}
The dynamical components $<k_T^2(n,Q^2)>_{i,{\rm evol}}$ of $<k_T^2(n,Q^2)>_i$ are given 
by Eq.~(\ref{ptn1}) with $f^{\prime}_i(n,Q^2)=f^{\prime}_{i,evol}(n,Q^2)$ corresponding 
to the solution of Eq.~(\ref{evold}) with the initial condition 
$f^{\prime}_i(n,Q_0^2)=0$.    
The solution of Eq.~(\ref{evold}) with the initial condition $\hat f^{\prime}(n,Q^2=Q_0^2)=0$ 
can be formally written as
\begin{equation}
\hat f^{\prime}_{evol}(n,Q^2)=-{1\over 4}\int_{Q_0^2}^{Q^2}dq^2 {\alpha_s(q^2)\over 
2\pi}\exp\{\hat P(n)[\xi(Q^2)-\xi(q^2)]\}\hat P^{\prime }(n)\hat f(n,q^2),
\label{solevol}
\end{equation}
where 
\begin{equation}
\xi(q^2)=
\int_{Q_0^2}^{q^2}{dq^{\prime 2}\over q^{\prime 2}}{\alpha_s(q^{\prime 2})\over 2\pi}.
\label{xia}
\end{equation}
This representation gives, after some algebra, the following explicit expressions 
for $f^{\prime}_{\rm S,evol}(n,Q^2)$ and $f^{\prime}_{\rm G,evol}(n,Q^2)$:
\begin{eqnarray}
&&f^{\prime}_{S,G,evol}(n,Q^2)= \nonumber \\
&&-{1\over 4}\{
\exp[\gamma^{+}(n)\xi(Q^2)][h_{S,G}^{++}(n)\eta(0,Q^2)+
h_{S,G}^{+-}(n)\eta(-\sqrt{\Delta(n)},Q^2)] \nonumber \\
&&+ \exp[\gamma^{-}(n)\xi(Q^2)][h_{S,G}^{--}(n)\eta(0,Q^2)+
h_{S,G}^{+-}(n)\eta(\sqrt{\Delta(n)},Q^2)]\},
\label{spsg}
\end{eqnarray}
where 
\begin{eqnarray}
\eta(r,Q^2)=\int_{Q_0^2}^{Q^2} dq^2 {\alpha_s(q^2)\over 2 \pi} \exp[r\xi(q^2)]
\label{eta}
\end{eqnarray}
 The functions  $\gamma^{mp}(n)$ are the eigenvalues of of the matrix $\hat P(n)$ and are 
given by   
\begin{equation}
\gamma^{\mp}(n)={P_{qq}(n)+P_{gg}(n)\mp \sqrt{\Delta(n)}\over 2},
\label{gampm}
\end{equation}
with 
\begin{equation}
\Delta(n)=[P_{qq}(n)-P_{gg}(n)]^2+4P_{qg}(n)P_{gq}(n).
\label{delta}
\end{equation}
The coefficients $h^{++}_{S,G}(n)$, $h^{+-}_{S,G}(n),h^{--}_{S,G}(n)$, and $h^{-+}_{S,G}(n)$ 
are expressed in terms of the input moments $f_{S,G}(n,Q_0^2)$ 
\begin{eqnarray}
h^{++}_{S}(n)&=&{1\over \sqrt{\Delta(n)}}\{[\gamma^{+}(n)-P_{gg}(n)][P^{\prime}_{qq}(n)
f_S^{+}(n,Q_0^2)+P^{\prime}_{qg}(n)
f_G^{+}(n,Q_0^2)] \nonumber \\ 
&+&P_{qg}(n)[P^{\prime}_{gq}(n)
f_S^{+}(n,Q_0^2)+P^{\prime}_{gg}(n)
f_G^{+}(n,Q_0^2)]\}
\label{hpps} \\
h^{++}_{G}(n)&=&{1\over \sqrt{\Delta(n)}}\{P_{gq}(n)[P^{\prime}_{qq}(n)
f_S^{+}(n,Q_0^2)+P^{\prime}_{qg}(n)
f_G^{+}(n,Q_0^2)] \nonumber \\
&+&[\gamma^{+}(n)-P_{qq}(n)][P^{\prime}_{gq}(n)
f_S^{+}(n,Q_0^2)+P^{\prime}_{gg}(n)
f_G^{+}(n,Q_0^2)]\} \\
\label{hppg}
h^{+-}_{S}(n)&=&{1\over \sqrt{\Delta(n)}}\{[\gamma^{+}(n)-P_{gg}(n)][P^{\prime}_{qq}(n)
f_S^{-}(n,Q_0^2)+P^{\prime}_{qg}(n)
f_G^{-}(n,Q_0^2)] \nonumber \\
&+&P_{qg}(n)[P^{\prime}_{gq}(n)
f_S^{-}(n,Q_0^2)+P^{\prime}_{gg}(n)
f_G^{-}(n,Q_0^2)]\} \\
\label{hpms}
h^{+-}_{G}(n)&=&{1\over \sqrt{\Delta(n)}}\{P_{gq}(n)[P^{\prime}_{qq}(n)
f_S^{-}(n,Q_0^2)+P^{\prime}_{qg}(n)
f_G^{-}(n,Q_0^2)] \nonumber \\
&+& [\gamma^{+}(n)-P_{qq}(n)][P^{\prime}_{gq}(n)
f_S^{-}(n,Q_0^2)+P^{\prime}_{gg}(n)
f_G^{-}(n,Q_0^2)]\} \\
\label{hpmg}
h^{--}_{S}(n)&=&-{1\over \sqrt{\Delta(n)}}\{[\gamma^{-}(n)-P_{gg}(n)][P^{\prime}_{qq}(n)
f_S^{-}(n,Q_0^2)+P^{\prime}_{qg}(n)
f_G^{-}(n,Q_0^2)] \nonumber \\
&+& P_{qg}(n)[P^{\prime}_{gq}(n)
f_S^{-}(n,Q_0^2)+P^{\prime}_{gg}(n)
f_G^{-}(n,Q_0^2)]\} \\
\label{hmms}
h^{--}_{G}(n)&=&-{1\over \sqrt{\Delta(n)}}\{P_{gq}(n)[P^{\prime}_{qq}(n)
f_S^{-}(n,Q_0^2)+P^{\prime}_{qg}(n)
f_G^{-}(n,Q_0^2)] \nonumber \\
&+&
[\gamma^{-}(n)-P_{qq}(n)][P^{\prime}_{gq}(n)
f_S^{-}(n,Q_0^2)+P^{\prime}_{gg}(n)
f_G^{-}(n,Q_0^2)]\} \\
\label{hmmg}
h^{-+}_{S}(n)&=&-{1\over \sqrt{\Delta(n)}}\{[\gamma^{-}(n)-P_{gg}(n)][P^{\prime}_{qq}(n)
f_S^{+}(n,Q_0^2)+P^{\prime}_{qg}(n)
f_G^{+}(n,Q_0^2)] \nonumber \\
&+& P_{qg}(n)[P^{\prime}_{gq}(n)
f_S^{+}(n,Q_0^2)+P^{\prime}_{gg}(n)
f_G^{+}(n,Q_0^2)]\} \\
\label{hmps}
h^{-+}_{G}(n)&=&-{1\over \sqrt{\Delta(n)}}\{P_{gq}(n)[P^{\prime}_{qq}(n)
f_S^{+}(n,Q_0^2)+P^{\prime}_{qg}(n)
f_G^{+}(n,Q_0^2)] \nonumber \\
&+& [\gamma^{-}(n)-P_{qq}(n)][P^{\prime}_{gq}(n)
f_S^{+}(n,Q_0^2)+P^{\prime}_{gg}(n)
f_G^{+}(n,Q_0^2)]\},
\label{hpmg2}
\end{eqnarray}
where 
\begin{eqnarray}
f_S^{\mp}(n,Q_0^2)&=&\mp{1\over \sqrt{\Delta}}\{[\gamma^{\mp}(n)-P_{gg}]f_S(n,Q_0^2)+
P_{qg}(n)f_G(n,Q_0^2)\}, \\
\label{fsmp0}
f_G^{\mp}(n,Q_0^2)&=&\mp{1\over \sqrt{\Delta}}\{[\gamma^{\mp}(n)-P_{qq}]f_G(n,Q_0^2)+
P_{gq}(n)f_S(n,Q_0^2)\}.
\label{fgpm0}
\end{eqnarray}
The functions  $f_{S,G}(n,Q^2)$ are given by:
\begin{equation}
 f_{S,G}(n,Q^2)=f_{S,G}^{+}(n,Q_0^2)\exp [\gamma^{+}(n)\xi(Q^2)]+ 
f_{S,G}^{-}(n,Q_0^2)exp [\gamma^{-}(n)\xi(Q^2)]
\label{fsgn}
\end{equation}

\end{document}